\documentclass[final,5p]{elsarticle}
\usepackage{eso-pic}
\usepackage{multirow}
\usepackage{subfigure}
\usepackage{url}
\usepackage{times}
\usepackage{array} 
\usepackage{tabulary}
\usepackage{colortbl}
\usepackage{calc}
\usepackage{amsmath}
\usepackage{setspace}

\usepackage[font=small, format=plain,labelfont=bf]{caption}

\journal{Pervasive and Mobile Computing}

\begin{document}

\begin{frontmatter}



\title{Mobile Multimedia Streaming Techniques: QoE and Energy Consumption Perspective}


\author[label1,label2]{Mohammad Ashraful Hoque}
\author[label1]{Matti Siekkinen}
\author[label1]{Jukka K. Nurminen}
\address[label1]{Aalto University School of Science, firstname.lastname@aalto.fi}
\author[label3]{Mika Aalto}
\address[label3]{Nokia Solutions and Networks, mika.aalto@nsn.fi}
\author[label2]{Sasu Tarkoma}
\address[label2]{University of Helsinki, firstname.lastname@cs.helsinki.fi}


\begin{abstract}
Multimedia streaming to mobile devices is challenging for two reasons. First, the way content is delivered to a client must ensure that the user does not experience a long initial playback delay or a distorted playback in the middle of a streaming session. Second, multimedia streaming applications are among the most energy hungry  applications in smartphones. The energy consumption mostly depends on the delivery techniques and on the power management techniques of wireless access technologies (Wi-Fi, 3G, and 4G). In order to provide insights on what kind of streaming techniques exist, how they work on different mobile platforms, their efforts in providing smooth quality of experience, and their impact on energy consumption of mobile phones, we did a large set of active measurements with several smartphones having both Wi-Fi and cellular network access. Our analysis reveals five different techniques to deliver the content to the video players. The selection of a technique depends on the mobile platform, device, player, quality, and service. The results from our traffic and power measurements allow us to conclude that none of the identified techniques is optimal because they take none of the following facts into account: access technology used, user behavior, and user preferences concerning data waste. We point out the technique with optimal playback buffer configuration, which provides the most attractive trade-offs in particular situations. 

\end{abstract}

\begin{keyword}
Performance analysis\sep measurement\sep power consumption\sep wireless multimedia \sep Quality of Experience (QoE). 

\end{keyword}

\end{frontmatter}




\section{Introduction}

Digital video content is increasingly consumed using mobile devices~\cite{cisco}. At the same time, the playback quality experienced by the user and the battery life of smartphones have become critical factors in user satisfaction. Consequently, it is essential that mobile video streaming not only provides a good viewing experience but also avoids excessive energy consumption.

Multimedia streaming services consider a number of challenges while sending content to the streaming clients for providing smooth playback, such as initial playback delay, clients with different kinds of connectivity, and the bandwidth variation between a server and a client~\cite{guole}. While consuming multimedia streaming content, energy consumption of smartphones is also considered as an important issue and consequently a significant number of research work focused on reducing energy consumption of mobile devices using streaming applications~\cite{hoque12survey}. The aforementioned streaming services have adopted various techniques to deliver video content to mobile users, such as encoding rate streaming, rate throttling, buffer adaptive streaming, fast caching, and rate adaptive streaming over HTTP. Encoding rate streaming is used to deliver content at the encoding rate. Throttling and fast aching send video content at a higher rate than the encoding rate. Buffer adaptive mechanisms work based on the playback buffer status of a client player. In this case, the client receives content from the server only when playback buffer drains to a specific lower threshold. Fast caching allows the player to download the whole content at the very beginning. Rate adaptive mechanisms adapt video quality according to the end-to-end bandwidth between a server and the client.

There has been work on analyzing the merits of these streaming techniques from the server performance point of view. For example, fast caching reduces start-up delay at the client and guards against bandwidth fluctuation, but it also consumes a lot of CPU and memory at the streaming server~\cite{guole}. Although most of the techniques are understood by research community, a thorough study of these streaming techniques is still required from the perspective of the mobile device and the user. Even though some studies have looked at the traffic pattern of video streaming services with Android, iOS devices, and desktop users~\cite{Rao:2011,Finamore:2011,Erman}, at present it is not well understood how the different techniques are chosen, how they compare to each other, and what are the optimal techniques to use in different contexts. Most importantly, the effect of these streaming techniques on user satisfaction on playback quality, Wi-Fi and cellular network usage, and on the energy consumption of mobile devices is yet to be fully uncovered. Such knowledge is imperative before one can design a streaming service that satisfies users demands in terms of quality of experience and battery life of their smartphones.



We actively captured traffic of more than five hundred video streaming sessions, from YouTube, Vimeo, Dailymotion and Netflix, via Wi-Fi, HSPA, and LTE. During those sessions we estimated the joining time. From the captured traffic we computed the playback buffer status. We also measured the energy consumption of smartphones during the streaming sessions. Our main observations are the following:
\vspace{4mm}
\begin{itemize}

\item In general, fast caching and throttling are applied by the
  server, whereas video players enforce encoding rate and buffer adaptive mechanism by exploiting TCP's flow control mechanism, hence, overriding the server selected mechanisms. In encoding rate streaming, the player unintentionally triggers TCP flow control because the player has too small playback buffer compared to the amount of content the server offers. The buffer adaptive mechanisms deliberately pause and resume download, and these techniques are applied only by the video players in Android phones. (Section~\ref{four})

\item Our analysis reveals that in smartphones different techniques are applied with little or no consensus: different techniques are used by different clients to access the same service in the same context. For example, Android devices use three different techniques for YouTube videos. The selection of those techniques depends on the quality of the video and the player. However, the strategy selection does not depend on the wireless interface being used for streaming and, thus, network operators do not play any role. (Section~\ref{four})

\item The joining time (a.k.a. initial playback delay) varies according to the wireless interface being used for streaming, the quality of the content, and the video service. The players experience shorter delay when streaming via Wi-Fi than HSPA or LTE. From the quality perspective, low quality videos are played with a shorter initial delay. Among the targeted video services, the Netflix players experience the longest delay. However, most of the streaming strategies are optimized for providing uninterrupted playback by allowing the players to keep a large amount of data in the playback buffer. (Section~\ref{qoe_streaming})
\item There is a large variation in playback energy consumption between different types of players and containers on the same device. The differences  are due to inefficient player implementation. However, the video quality (resolution) does not seem to have a large impact on energy consumption. (Section~\ref{five_one})

\item When the user views the entire video clip, fast caching and throttling are the most optimized techniques for providing uninterrupted playback at the client. At the same time, they are the most energy efficient. 
If the user is likely to interrupt the video viewing, buffer adaptive streaming is more attractive as the player generates ON-OFF traffic pattern and less energy is consumed for wireless communication during an OFF period. However, the ON period duration should be adjusted to match fast start period in order to avoid server rate throttling. Similarly, the duration of the OFF period should also be optimized so that the player does not suffer from playback buffer starvation. However, none of the identified techniques alone is optimal because they do not adapt to the wireless access technology, user behavior, and preferences. (Section~\ref{five_three})

\end{itemize}

We structure our paper as follows. In the next section, we briefly describe the energy consumption characteristics of wireless communication in smartphones, explain the characteristics of mobile video streaming. In Section~\ref{three}, we describe our measurement and data collection methodology. In Section~\ref{four}, we investigate the different streaming techniques. Section~\ref{qoe_streaming} examines the effort of the streaming techniques in providing uninterrupted playback. Section~\ref{five} is devoted to presenting the results from the energy consumption measurements. In section~\ref{tradeoff}, we discuss the tradeoff between energy savings and potential playback buffer underrun. Finally, we contrast our work with earlier research in Section~\ref{seven} before concluding the paper.

\section{Background}

Smartphones allow users to access Internet via Wi-Fi and mobile broadband access. Mobile broadband experience is enabled by the latest 3G and 4G technologies such as EV-DO, HSPA, and LTE. The most widely deployed mobile broadband technology is currently HSPA, while LTE is the fastest ever growing cellular and mobile broadband technology. In this section, we first review the power consumption characteristics of Wi-Fi and cellular interfaces that we use in this study. Then, we explain the characteristics of mobile streaming services and the metrics to assess the quality of experience of the users.

\subsection{Power Saving Mechanisms for Wireless Network Interfaces}
\label{sec:powersave}

\subsubsection{Wi-Fi} Smartphones implement 802.11 Power Saving Mechanism (PSM) to manage the power consumption of Wi-Fi. There are four states; transmit, receive, idle and sleep. PSM allows the interface to be in sleep when there is not data activity. However, the client periodically powers up the interface to receive a traffic indication map (TIM) frame from the access point (AP). This interval is usually 100ms and also called listen interval. The TIM frame tells a mobile client whether the AP has some buffered data for the mobile device or not. If the AP has data for the client, the client sends PS-Poll frame in return to receive the buffered data. Otherwise, the client goes back to sleep. Modern devices usually implement a timer which keeps the interface in idle state for a few hundred milliseconds after the transmission or reception of packets, which improves especially the performance of short TCP connections. This is also known as PSM adaptive~\cite{TanGCZ07}.

\label{two}
\begin{figure}[t]
\centering
\includegraphics[width=0.8\linewidth, height=0.5\linewidth]{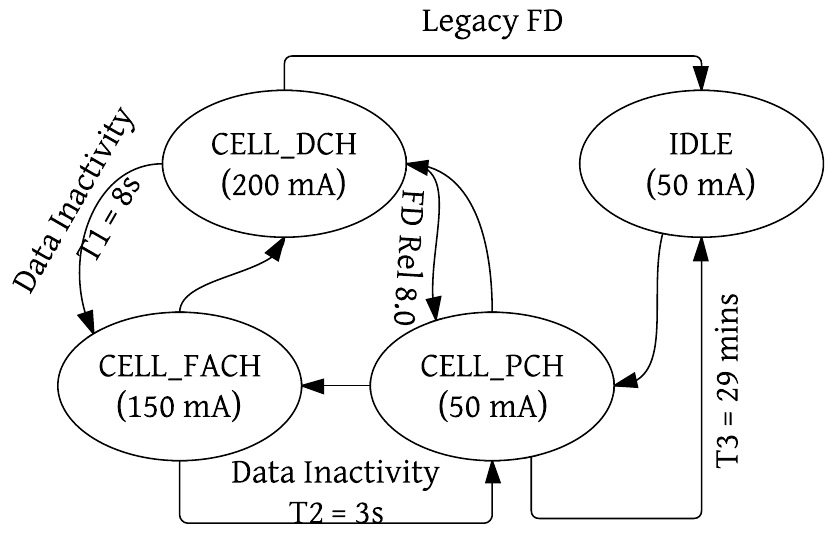}
\caption{WCDMA/HSPA RRC states with typical values of the inactivity timers and power consumption.}
\label{fig:hspa_states}
\end{figure}

\subsubsection{WCDMA/HSPA}
3GPP standards specify the efficient usage of the radio resources considering the mobility and power consumption of smartphones via a resource control protocol (RRC). Figure~\ref{fig:hspa_states} shows that there are a number of states and inactivity timers in 3GPP RRC protocol. These timers ensure that if a certain resource is not utilized for a certain period of time in a particular state, the resource must be released. For example, high volume data transmission happens in CELL\_DCH state and small packet transmission is possible in CELL\_FACH state. A mobile device switches from CELL\_DCH to CELL\_FACH in absence of data activity for a period of T1 seconds. These timers have static values and they are operator controlled. If the mobile device and network both support Rel 8.0 Fast Dormancy (FD)~\cite{fastdorm}, CELL\_DCH$\rightarrow$ CELL\_PCH transition happens. For non standard FD, the transition is CELL\_DCH$\rightarrow$ IDLE (Figure~\ref{fig:hspa_states}) which releases the RRC connection altogether.

RRC protocol has a large impact on the energy spending of smartphones. Figure~\ref{fig:hspa_states} also shows that average current consumption in CELL\_DCH is 200mA, in CELL\_FACH is 150mA, and in CELL\_PCH is 50mA approximately. The potential consequence especially with long inactivity timers is high power consumption at the mobile device. To learn more about different cellular network configurations and their effect on energy consumption, readers can follow~\cite{siekkinen2013movid}.

\begin{figure}[!t]
\centerline{\includegraphics[scale=0.53]{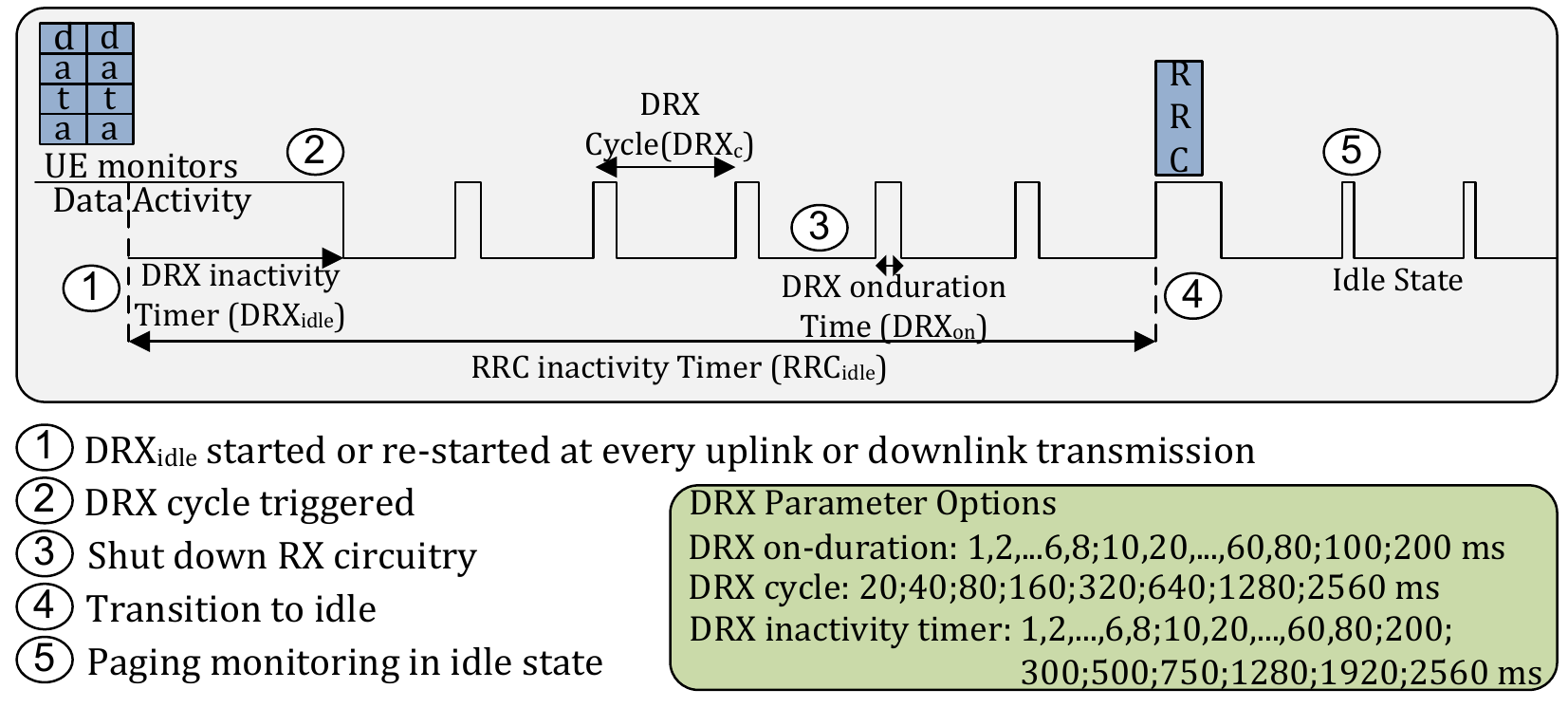}}
\caption{LTE DRX Cycles and timers.}\label{fig:lte_drx_rrc}
\end{figure}

\subsubsection{LTE}
The radio resource control protocol for LTE specifies only two states; RRC\_IDLE and RRC\_CONNECTED. Similar to the HSPA RRC protocol, an inactivity timer (RRC$_{idle}$) controls the connected to idle state transition. LTE includes a discontinuous transmission and reception (DTX/DRX) mechanism that enables a mobile device to consume low power even being in the RRC\_CONNECTED state. DRX in the connected state is also called connected mode DRX or cDRX, and the associated inactivity timer is DRX$_{idle}$. Figure \ref{fig:lte_drx_rrc} shows that if there is no data activity for DRX$_{idle}$ time, then a DRX cycle, DRX$_c$, is initiated. The length of a such cycle can vary from 20ms to  few seconds. The device checks data activity during the on period, DRX$_{on}$, of the cycle. If the data inactivity continues for a long time, RRC$_{idle}$, the network commands the device to switch from RRC\_CONNECTED to RRC\_IDLE state. Then the device enters in the paging monitoring mode in the IDLE state.

\subsection{Mobile Video Streaming}

\begin{table}[!t]
  \begin{center}
  {\footnotesize
    \begin{tabular}{|p{40mm}|p{40mm}|}
          \hline
 Streaming Services & YouTube, Vimeo, Dailymotion, and Netflix\\\hline 
 Players & Native Application, Flash, and HTML5\\\hline
 Video Quality & LD (240p), SD (270-480p), HD (720-1080p)\\\hline
 Containers & 3GPP, MP4, WebM, X-FLV, ismv\\\hline
      \end{tabular}}
                 \caption{Streaming services, the players used by  the clients for playback, the quality of the content and the containers to deliver the content.}
          \label{tabvs}
   \end{center}
\end{table}

Today mobile streaming services deliver content using HTTP over TCP. Smartphone users can access these services using either a native app or a browser. The browser may load a Flash, HTML5 or Microsoft Silverlight player. The quality of the video played is often denoted with a p-notation, such as 240p, which refers to the resolution of the video. 240p usually refers to 360x240 resolution. Different services use also low, standard, and high definition (LD, SD, HD) notations but the resolutions that each one refers to varies between services. Therefore, we define 240p videos as LD, 270-480p videos as SD and 720-1080p or higher resolution videos as HD. MP4, WebM, and X-FLV are the default containers for the players. The native apps of YouTube, Dailymotion and Vimeo also play MP4 and 3GPP videos. Netflix players play ismv videos. WebM and X-FLV are the default containers for the HTML5, and Flash player respectively. Table~\ref{tabvs} shows the examples of examples of different video services, the types of video players, video qualities, and containers. 


\subsection{Quality of Experience}

The quality of streaming perceived by a user is influenced by the network condition, content quality (e.g. HD or SD), user's preference on the content, and the context in which the user is viewing a video. The network condition translates to network congestion caused by the bottleneck point in between a streaming client and the server. This network congestion is evidenced by the reduced available bandwidth and packet loss. The impact is realised by the user as long initial playback delay and pauses or playback starvation during playback. In wireless networks, the bottleneck situation can arise when multiple users share the common resources and the throughput per user is reduced so much that user experience is degraded. The bottleneck also can be caused by the radio conditions, i.e. in cell edge the available bit rate is much lower than peak HSPA/LTE bit rates even in an empty cell. The state transition of the WNIs can introduce additional delays.

For dealing with various network conditions, video services apply a number of strategies; i) Encoding rate streaming, ii) Throttling, iii) Buffer Adaptive Streaming, iv) Rate Adaptive Streaming, and v) Fast Caching. A common feature of all streaming services is an initial buffering of multimedia content at the client. This initial buffering is also referred to as Fast Start. The name comes from the fact that a player downloads content using all the available bandwidth. Fast caching is similar to Fast Start, the only difference is that fast caching lasts longer until the whole content is downloaded. These techniques are used by the services for constant bit rate streaming, except rate adaptive streaming. The most prevalent forms of rate adaptive streaming are HTTP live streaming (HLS), Microsoft Smooth Streaming (MSS).


\begin{figure}[t]
\centering
\includegraphics[width=0.9\linewidth, height=0.35\linewidth]{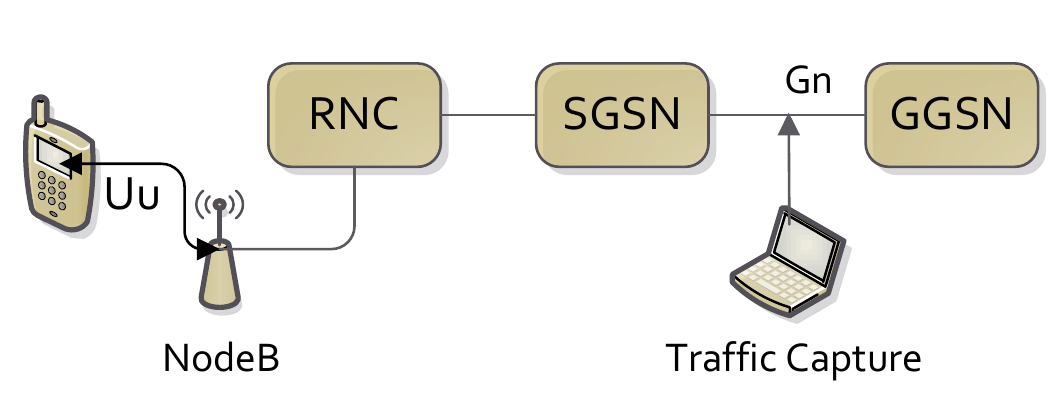}
\caption{Capturing traffic at the Gn interface between SGSN and GGSN in the test HSPA Network.}
\label{fig:hspacapture}
\end{figure}
\begin{figure}[t]
\centering
\includegraphics[width=0.9\linewidth, height=0.4\linewidth]{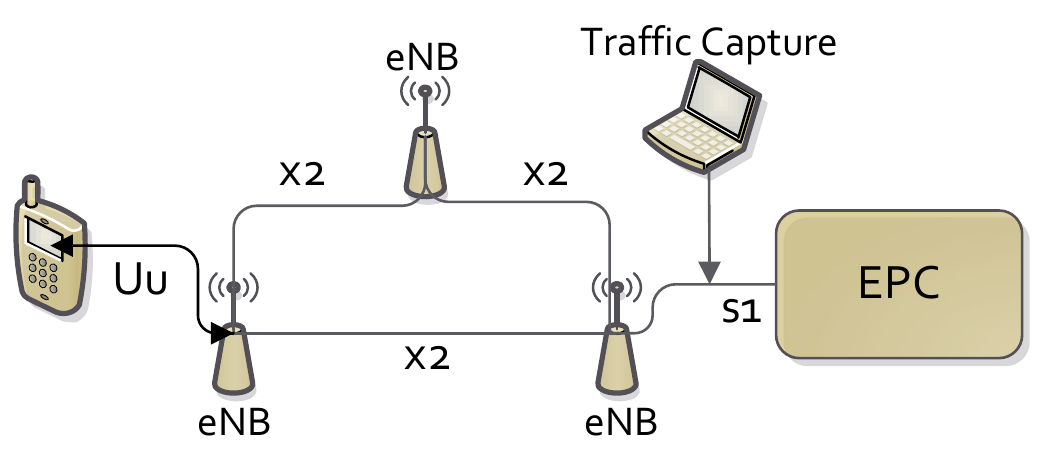}
\caption{Capturing traffic at the S1 interface between the eNB (Base Station) and EPC (Evolved Packet Core) in the LTE network.}
\label{fig:ltecapture}
\end{figure}

\section{Measurement and Data Collection}
\label{three}

\subsection{Properties of the multimedia content}
Compared with our earlier work~\cite{hoque2013wowmom}, we excluded previous results for Meego, Symbian, and WP7.5 platforms. We included three latest smartphones; iPhone5, Galaxy S3 LTE (GS3 LTE) and Lumia825. All the video services, YouTube, Dailymotion, Vimeo and Netflix, have the native applications for the target mobile platforms. The desktop edition of YouTube was used only in the Android platforms as it provides the opportunity to use both Flash and HTML5 players. Our target video services, players, and smartphones are listed in Table~\ref{tab:mobile}. Whenever available for the particular smartphone and player, we streamed videos of multiple qualities that range from LD to HD. The average duration of the videos was 10 minutes.

\begin{table}[!t]
  \begin{center}
  {\footnotesize
    \begin{tabular}{|p{40mm}|p{40mm}|}
          \hline
 Config Name & Parameters\\\hline 
 noDRX & RRC\_${idle}$=10 s\\\hline 
DRX$_{80ms}$ & RRC$_{idle}$=10 s, DRX$_{cycle}$=80 ms, DRX$_{on}$=10 ms, \\\hline 
DRX$_{160ms}$ & RRC$_{idle}$=10 s, DRX$_{cycle}$=160 ms, DRX$_{on}$=10 ms,\\\hline 
DRX$_{640ms}$ & RRC$_{idle}$=10 s, DRX$_{cycle}$=640 ms, DRX$_{on}$=10 ms,\\\hline 

      \end{tabular}}
             \caption{LTE network configurations.}

          \label{tab:drx_profiles}
   \end{center}
\end{table}

\begin{table*}[!t]
  \begin{center}
  {\footnotesize
    \begin{tabular}{|p{16mm}|p{20mm}|p{20mm}|p{19mm}|p{19mm}|p{18mm}|}
      \hline
 		&  \textbf{iPhone4S}&  \textbf{iPhone5} & \multicolumn{2}{c|}{\textbf{Galaxy S3/\break Galaxy S3 LTE\break (Android-4.0.4)}}& \textbf{Lumia825}\\
 		 &  \textbf{iOS 5.0} & \textbf{(iOS 7.0)} &\multicolumn{2}{c|}{}  & \textbf{(WP8)} \\\hline
 		\textbf{YouTube}\break Streaming&\textbf{(App)} Throttling\break Factor=2.0&\textbf{(App)} Throttling\break Factor=1.25 & \textbf{(Flash)} \break Encoding rate(HD),\break Throttling($<$HD)\break Factor=1.25&\textbf{(App\& HTML5)} ON-OFF-M &\textsf{(App)}\break Fast Caching \\\cline{1-6}
 		Quality & LD(240p), SD(360p), HD(720p)& LD(240p), SD(360p), HD(720p)&LD(240p),\break SD(360,480p),\break HD(720,1080p) & LD(240p),\break SD(360,480p),\break HD(720p) & SD(270p), HD(720p)\\\cline{1-6}
 		Container&MP4(360,720p)&MP4(360,720p)\break 3GPP(240p)&  XFLV & MP4($>$240p)\break WebM($>$240p)\break 3GPP(270p)&MP4(720p)\break 3GPP(270p)  \\\hline\hline
		\textbf{Vimeo}\break Streaming&\textbf{(App)}\break HLS\break Chunk Size=10s&\textbf{(App)}\break ON-OFF-M& \multicolumn{2}{c|}{\parbox[t]{15mm}{\textbf{(App)}\\ ON-OFF-S}} & \textbf{(App)}\break Fast Caching\\\cline{1-6}
		Quality&*&SD(270,480p), HD(720p)&\multicolumn{2}{c|}{SD(270p), HD(720p)}&HD(720p)\\\cline{1-6}
 		Container& MP4& MP4&\multicolumn{2}{c|}{MP4}&MP4\\\hline\hline
 		\textbf{Dailymotion}\break Streaming&\textbf{(App)}\break Throttling\break Factor=1.25&\textbf{(App)}\break HSL\break Chunk Size=10s &\multicolumn{2}{c|}{\parbox[t]{25mm}{\textbf{(App)}\\ Fast Caching(288p),\\  ON-OFF-S($>$288p)}} & \textbf{(App)}\break Throttling\break Factor=1.25\\\cline{1-6}
 		Quality&LD(240)&*&\multicolumn{2}{c|}{SD(288,480p),\break HD(720p)}&SD(288p)\\\cline{1-6}
 		Container& MP4& MP4& \multicolumn{2}{c|}{MP4} & MP4  \\\hline\hline 
 		
 		\textbf{Netflix}\break Streaming&\textbf{(App)}\break HLS\break Chunk Size=10s &\textbf{(App)}\break HLS\break Chunk Size=10s & \multicolumn{2}{c|}{\parbox[t]{25mm}{\textbf{(App)}\\ON-OFF-S}} & \textbf{(App)}\break MSS\break Chunk Size=4s\\\cline{1-6}
 		Quality&*&*&\multicolumn{2}{c|}{HD(720p)}&*\\\cline{1-6}
 		Container& isma, ismv& isma, ismv& \multicolumn{2}{c|}{MP4} & isma, ismv  \\\hline 
      \end{tabular}}
             \caption{Streaming techniques for popular video streaming services to mobile phones of three major platforms. The selection of a streaming technique does not depend on the wireless interface being used for, rather depends on the player, video quality, device and the video service provider.}

          \label{tab:mobile}
   \end{center}
\end{table*}

\subsection{Network Setup}
We watched videos from the video services in the smartphones via Wi-Fi, HSPA, and LTE. In the case of Wi-Fi, a 802.11 b/g access point was used. The access point was connected to the Internet via 100 Mbps Ethernet. AirPcap\footnote{AirPcap:www.cacetech.com/documents/AirPcap\%20Nx\%20Datasheet.pdf} was used to capture the Wi-Fi traffic. HSPA network measurements were conducted in the Nokia Solutions and Networks test networks. The network parameters, i.e. states and inactivity timers,  were configured according to the vendor recommendation. The values of the inactivity timers were from few seconds to few minutes; T1=8s,T2=3s,T3=29min. The CELL\_PCH state was enabled in the network. We captured traffic of the streaming clients at the Gn interface between SGSGN and GGSN (see Figure~\ref{fig:hspacapture}). The LTE measurements were conducted with connected mode DRX enabled in the network. Traffic capture is taken at the S1 interface between the eNB and EPC. We measured power consumption with three sets of DRX profiles. The DRX profiles are described in Table~\ref{tab:drx_profiles}.


\subsection{Power Measurement}
We used Monsoon\footnote{Monsoon Power Monitor : www.msoon.com} and another custom power monitor for measuring the energy consumption of the smartphones during multimedia streaming. We removed the battery of most of the mobile phones and powered them using the measurement devices. Only the iPhones get power from the battery. All the devices were in automatic brightness settings during the power measurements.

\section{Streaming Techniques}
\label{four}
From traffic traces we inferred manually the type of streaming technique used for each of
the different combinations of device, service, stream quality, player
type, and access network type. These findings are summarized in Table \ref{tab:mobile} and discussed below.

\begin{figure}[t]
\centering
\includegraphics[width=0.8\linewidth, height=0.45\linewidth]{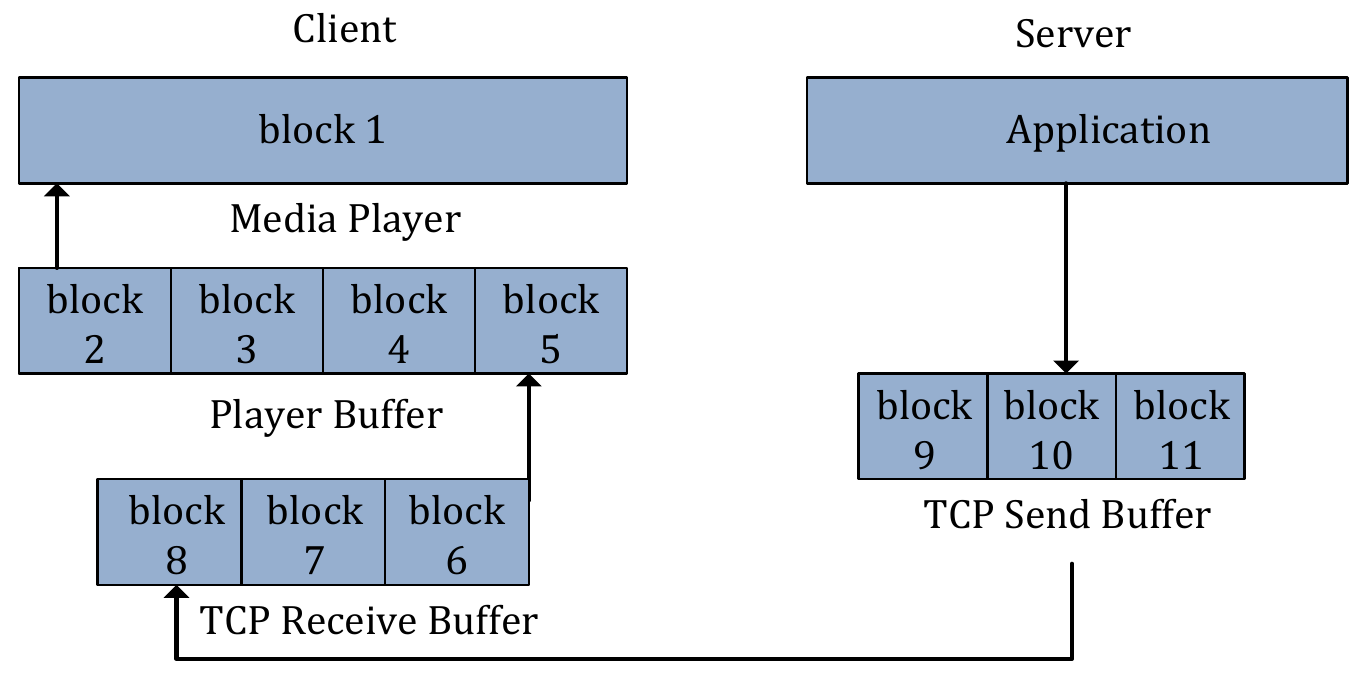}
\caption{Interaction between playback buffer and TCP receive buffer for encoding rate streaming.}
\label{fig:tcpbuffer}
\vspace{-5mm}
\end{figure}

\subsection{Encoding Rate Streaming}
\label{four_one}

Encoding rate technique is exclusively applied by the streaming clients. The server sends content using fast caching and the player has a small playback buffer. Therefore, the playback buffer and TCP receive buffer become full at the very beginning.  Since the player decodes content at the encoding rate, the same amount of buffer is freed from the playback buffer and also from the TCP receive buffer. The client again can receive the same amount of content from the server. The mechanism is illustrated in Figure~\ref{fig:tcpbuffer}. From Table~\ref{tab:mobile}, we can see that the Flash player in Android devices receives HD videos from YouTube at the encoding rate.

\subsection{Throttling}
\label{four_two}

Throttling is a server-side streaming technique. In this case, the server sends content at a limited constant rate, which is higher than the encoding rate. Therefore, the content is downloaded at the client at a faster pace than the encoding rate. The multiple of the encoding rate is referred to as the throttle factor. The throttling factor can vary depending on the video service or even on the player type for the same service. For instance, the native YouTube application receives content at a throttled factor of 2.0 in iPhone4S, whereas the Dailymotion application receives at a factor of 1.25. The Flash player in Android devices and the native app in iPhone5 specify the throttling factor in the request URL (e.g., \textsf{algorithm = throttle-factor} and \textsf{factor = 1.25}) or a service specific default throttle factor is used.

\begin{figure}[t]
  \begin{center}
    \subfigure[CDF of the chunk sizes YouTube.]{\label{fig:cdf_bursts}\includegraphics[width=0.49\linewidth, height=0.55\linewidth]{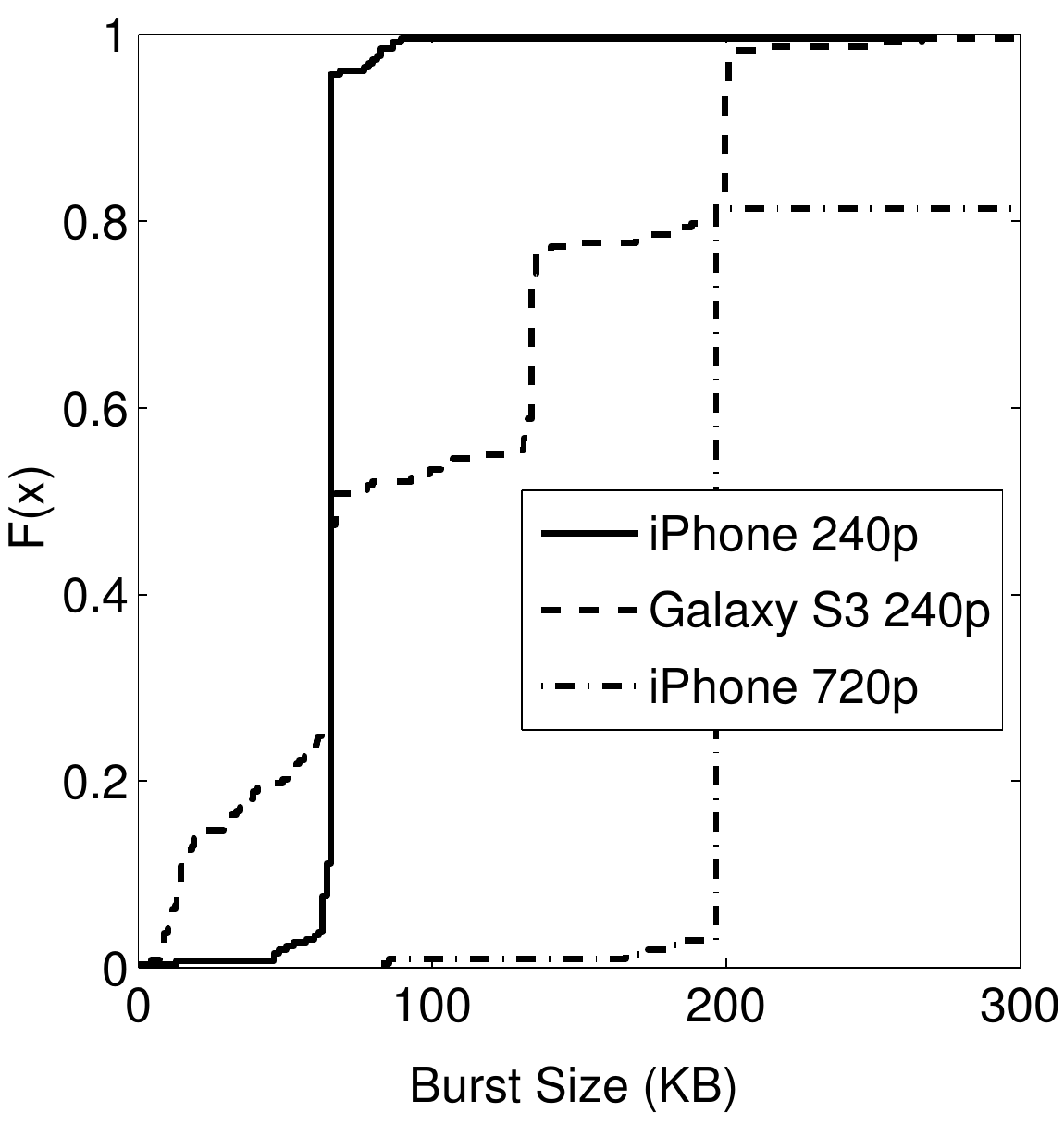}}
      \subfigure[CDF of the chunk intervals.]{\label{fig:cdf_burstival}\includegraphics[width=0.49\linewidth, height=0.55\linewidth]{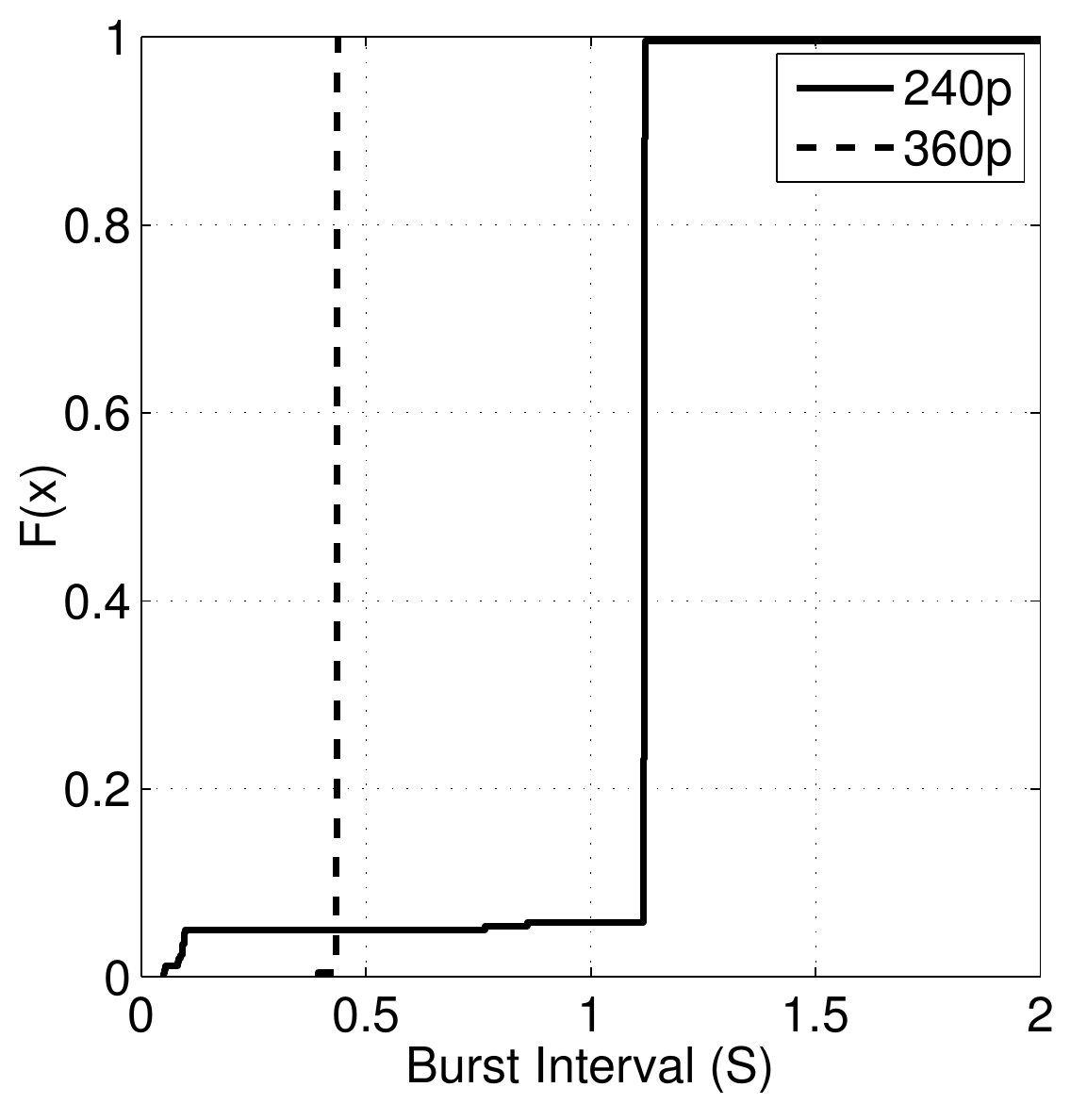}}
  \caption{YouTube server sends content in small chunks and periodic manner when throttling the sending rate.}
    \label{fig:throttle_burst}
 \end{center}
\end{figure}

\subsubsection{Single TCP connection}
In general, throttling is carried over a single TCP connection and the data is sent in small chunks. Figure~\ref{fig:cdf_bursts} shows that in the YouTube player in iPhone receives a LD video in 64KB chunks. This observation is similar to those explored in~\cite{Alcock2011} and~\cite{Rao:2011} for YouTube. The chunk size increases to 192KB when receiving the same video of HD quality. We observed variable chunk sizes when streaming to Samsung Galaxy S3 (see Figure~\ref{fig:cdf_bursts}). However, these chunks are sent by the streaming servers at some periodic intervals to the streaming clients. The interval increases as the encoding rate or quality of the video decreases. Figure~\ref{fig:cdf_burstival} shows that the chunks are separated by few hundred milliseconds to 1.2s. This burstiness is independent of the wireless interface being used at the client to receive the content. Nevertheless, this kind of burstiness was absent in Dailymotion and Vimeo traffic.

\subsubsection{Multiple TCP connections}
\label{sec:iphone_cnx}

In iPhone4S, the YouTube application uses a significant number of TCP connections to receive HD quality videos. In an example video session, we found that the player downloads a HD video in 66 connections. The player maintains a 25MB size playback buffer. At the beginning, the player receives content at the throttled rate. Since the playback continues at the encoding rate, there is always some extra content in the buffer. Therefore, this playback buffer becomes full at some point and the player closes the existing TCP connection. Whenever some buffer is freed, the player initiates another HTTP partial content request over TCP.


In this way, the player actually receives more data from the server than the actual size of the content. Finamore et al.~\cite{Finamore:2011} also reported similar observation. From traffic traces, we identified that a YouTube server always sends media content from the beginning of a key frame for any partial content request. The reason is that the player is unable to keep track of the ending position of the current key frame or the beginning of the next key frame. Therefore, it may terminate the connection when receiving a key frame. In addition, the player must support the forward and backward seeking during playback. Subsequently, each time the player requests content from the beginning of a key frame, which it has received partially for the previous request. As a result, the player wastes all the data of the partially received key frame. From traffic traces, we calculated that the player received 160MB data in total for for a 76MB video. 

\begin{figure}[!t]
  \begin{center}
    \subfigure[Buffer adaptive streaming over a single TCP connection activates TCP flow control during an OFF period.]{\label{fig:nexus_tcpflow}\includegraphics[width=0.7\linewidth, height=0.28\textwidth]{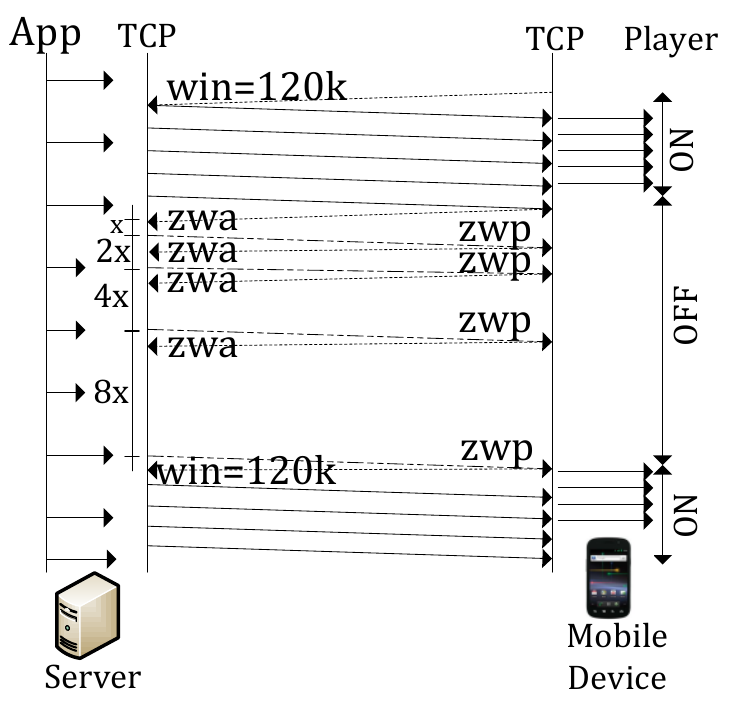}}\\
      \subfigure[The growth of the TCP persist timer at the streaming servers during an OFF period.]{\label{fig:tcp_persist}\includegraphics[width=0.7\linewidth, height=0.25\textwidth]{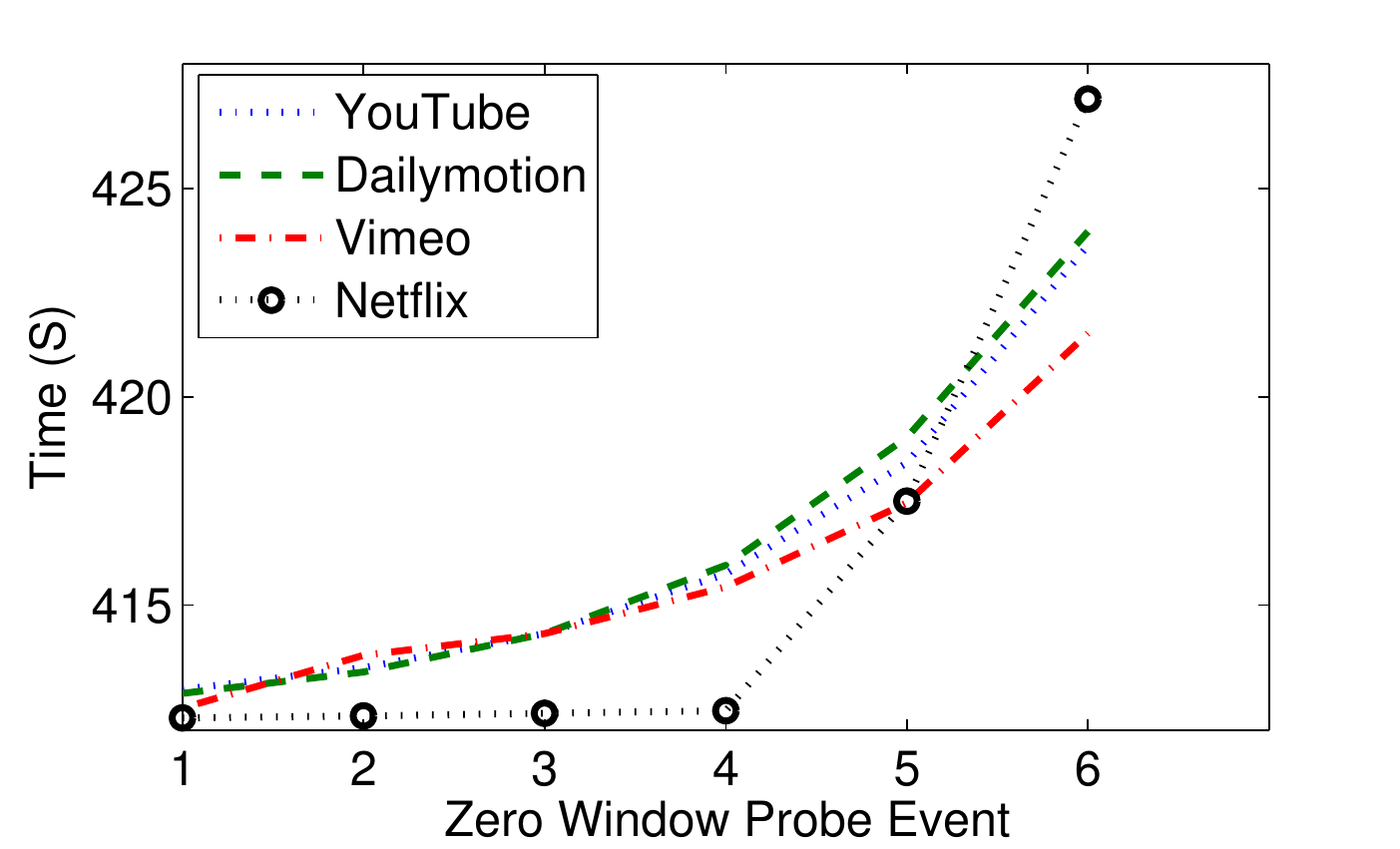}}
  \caption{ON-OFF-S mechanism and interaction with TCP flow control.}
    \label{fig:onoff_tcp}
 \end{center}
\end{figure}

\subsection{Buffer Adaptive Streaming}
\label{four_three}

Buffer adaptive techniques represent smart player implementation. The players maintain two thresholds of buffer level: a lower and an upper. During a streaming session, the player stops downloading content when the playback buffer is filled to the upper threshold value and it resumes downloading when the buffer drains to the lower threshold. The video players apply buffer adaptation in two different ways and generate ON-OFF traffic pattern. Some video players apply the buffer adaptation over a single TCP connection. We refer this kind as ON-OFF-S. The others use multiple TCP connections and we refer as ON-OFF-M.

\subsubsection{Single Persistent TCP Connection (ON-OFF-S)}
\label{on-off-s}
The native applications of Dailymotion, Vimeo, and Netflix video services apply buffer adaptation over a single TCP connection in the Android devices (see Table~\ref{tab:mobile}). The players stop reading from the TCP socket and an OFF period begins. Figure~\ref{fig:nexus_tcpflow} illustrates that TCP flow control packets are exchanged during an OFF period.

\begin{figure}[t]
\centering
\includegraphics[width=0.75\linewidth, height=0.65\linewidth]{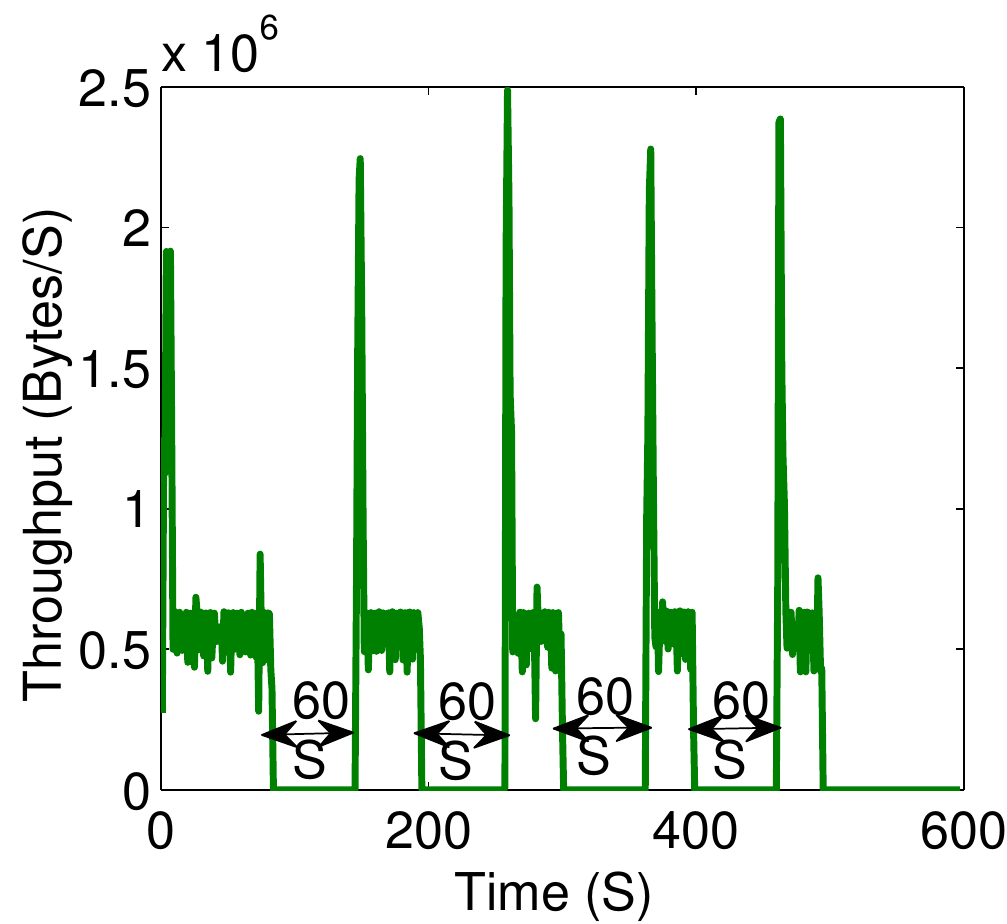}
\caption{The YouTube player in Galaxy S3 downloads a video by initiating multiple TCP connections.}
\label{fig:ON-OFF}
\end{figure}

The duration of an OFF period can be very long. The older Android devices (e.g., Samsung Nexus S) use an upper threshold of 5MB~\cite{hoque2013wowmom}. Therefore, the duration of the OFF period is almost equivalent to the $\frac{5 MB}{Encodingrate}$s. On the other hand, in latest devices the duration is $\frac{20 MB}{Encodingrate}$s. However, from traffic traces we found that the TCP persist timer at the server grows only to maximum 5s. The reason is that the players intentionally reset the persist timer after every 16s by receiving 64KB data from the server. This behavior was absent in the case of Netflix. Figure~\ref{fig:tcp_persist} shows how the TCP persist timer values grow at the servers of different video streaming services. In the case of Netflix, the OFF period is always 30s and the persist timer increases to maximum 10s. Later in Section~\ref{five_three}, we will see how the TCP flow control messages and TCP persist timer affects the power consumption of smartphones.

\subsubsection{Non-persistent TCP connections (ON-OFF-M)}

Only the native app and HTML5 player for YouTube in Android devices use multiple TCP connections for buffer adaptation. The players maintain dynamic lower and upper thresholds of playback buffer. When the playback buffer is filled to the upper threshold, the player closes the TCP connection and an OFF period begins. The ON period begins after a fixed 60s OFF period (see Figure~\ref{fig:ON-OFF}). The recent version of Vimeo player in iPhone5 also uses multiple TCP connections. Unlike the YouTube player, the Vimeo player downloads 30MB during the Fast Start and downloads rest of the content in 5MB chunks. Therefore, the duration of an OFF period is equal to $\frac{5 MB}{Encodingrate}$s.

\begin{figure}[!t]
  \begin{center}
    \subfigure[The Vimeo player in iPhone4S, using HLS, discards content of low quality from the playback buffer upon switching to a higher quality.]{\label{fig:vimeo_variation}
\includegraphics[scale=0.45]{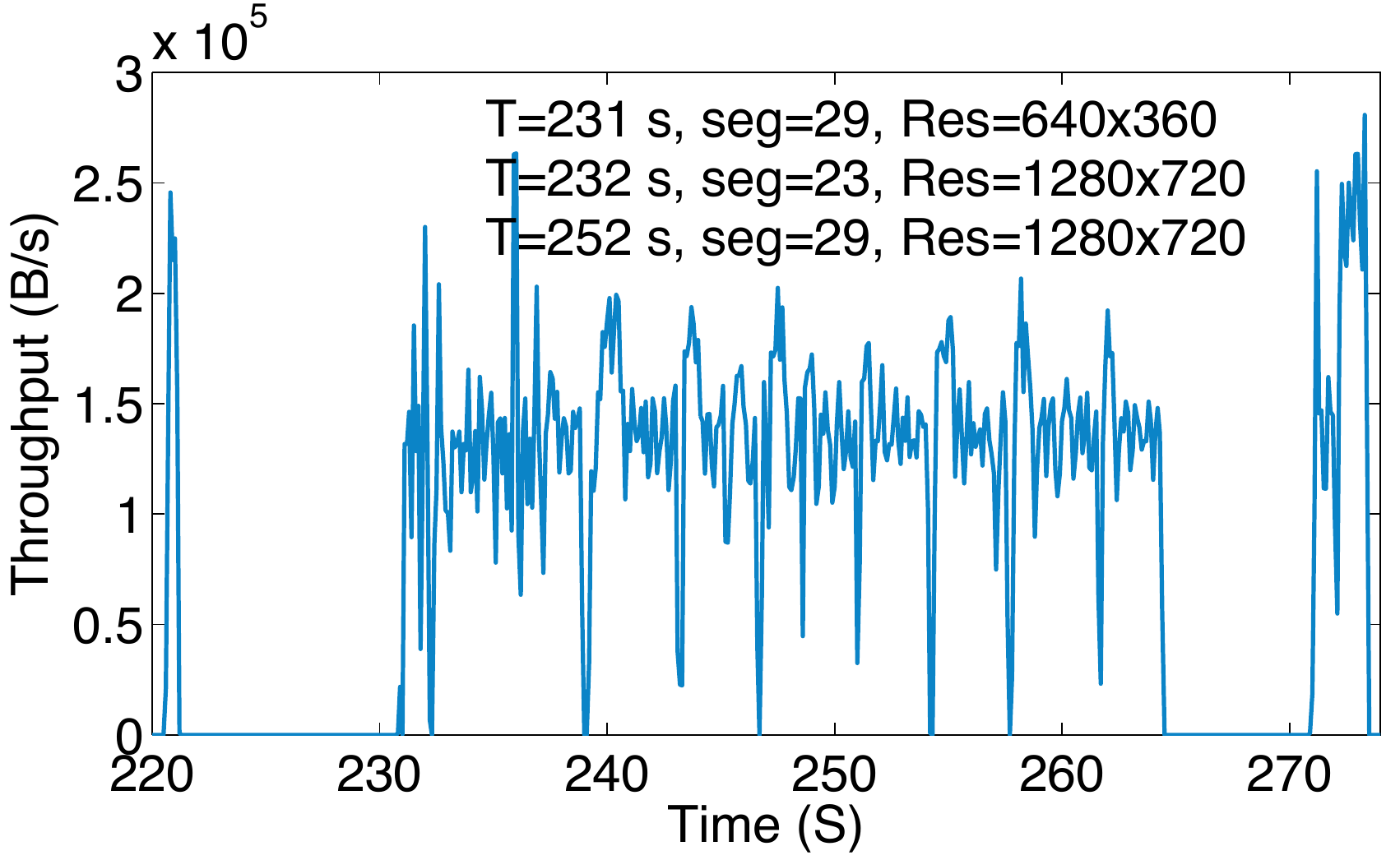}}
    \subfigure[The Netflix player in iPhone5, using HLS, downloads audio and video chunks asynchronously.]{\label{fig:netflix_variation}
\includegraphics[scale=0.62]{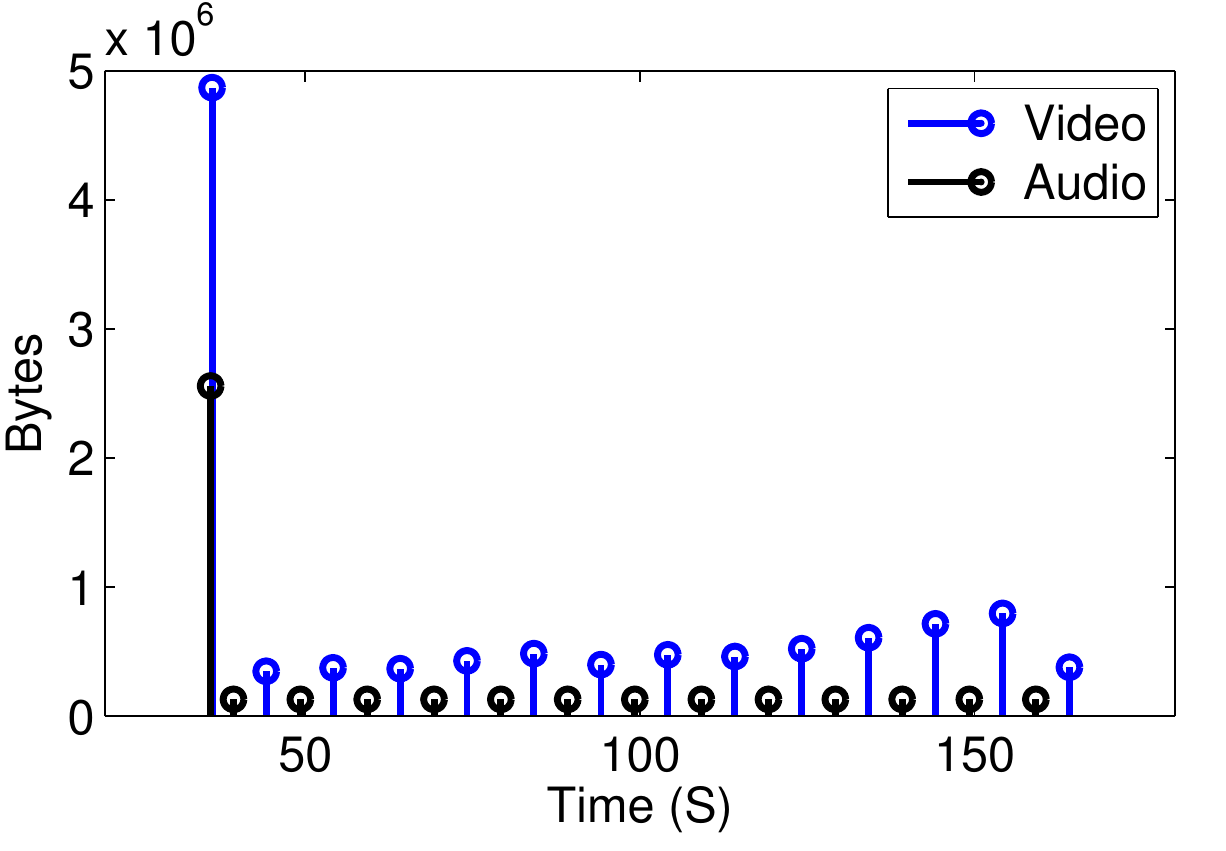}}
    \caption{Joining observed for the video services and when streaming via wireless network interfaces.}
    \label{fig:hsl_adaptive}
 \end{center}
\end{figure}

\subsection{Fast Caching}
Fast caching refers to downloading the whole content in one go at the very beginning of the streaming using the maximum available bandwidth. The players continue playback and at the same time maintains very large playback buffer. YouTube Flash player uses \texttt{ratebypass=yes} parameter in the HTTP request to deactivate any rate control at the server side. For example, the YouTube player of Lumia825 downloaded a 10-minute long $720p$ video within 120s via LTE or HSPA in our experiments. Lumia825 also receives video content from Vimeo at possible maximum rate.

\subsection{Rate Adaptive Streaming}
\label{four_five}
The streaming techniques we discussed so far are for streaming constant quality content during a streaming session. The players or the servers cannot change the quality on the fly, unless the user interrupts the playback. On the other hand, Dynamic Adaptive Streaming over HTTP (DASH~\cite{Stockhammer:2011})-like rate adaptive mechanisms are able to change the quality on the fly for adapting with bandwidth fluctuations. The quality switching algorithms are implemented in the client players. A player estimates the bandwidth continuously and transitions to a lower or to a higher quality stream if the bandwidth permits. We identified two kinds of rate adaptive streaming; (i) HTTP Live Streaming (HLS) and (ii) Microsoft Smooth Streaming (MSS).

\subsubsection{HTTP Live Streaming}
The Netflix and Vimeo players in iPhone4S, and the Dailymotion player in iPhone5 use HTTP Live Streaming and downloads content in 10s chunks. At the beginning, a player receives the media description files, which contain the chunk duration, encoding rates and the bandwidth requirements for the chunk download. The player begins by downloading seven 10s chunks of the SD quality. After that, the player downloads chunks after every ten seconds. In this way, the player always keeps 60s playback content in the buffer when streaming via Wi-Fi. In case of transitioning to a higher quality, the player discards the downloaded lower quality content in order to provide instant response to the quality change to the user. One streaming scenario via Wi-Fi is illustrated in Figure~\ref{fig:vimeo_variation}, where the player switches from a SD to HD quality at 232s and downloads from 23rd to 29th segments of HD quality. In the case of HSPA, the player wastes 20s content. This observation can change with bandwidth variation.   

Similarly, the Netflix player uses HLS in iPhones. However, the Netflix downloads the audio and video chunks separately, where the chunks are of 10s. From multiple traces, we verified that the audio and video chunk downloading are not synchronized. Figure~\ref{fig:netflix_variation} shows that after the Fast Start phase, the interval between an audio and a video chunk is approximately five seconds. There were also some cases where an audio chunk appears very close to the next video chunk. Another interesting observation is that the server specifies it's TCP parameters in the HTTP response header, as for example ~\textsf{X-TCP-Info:rtt=11625;snd\_cwnd=217201;rcv\_wnd=1049800}. The reason is likely that the streaming server lets the client player to calculate the bandwidth and to decide the quality accordingly.

\subsubsection{Microsoft Smooth Streaming}

The Netflix player in Lumia825 uses Microsoft's smooth streaming. The player receives video content in 4s chunks over a single TCP connection. The same connection is used to receive audio content chunks also. However, the audio chunks are received after every sixteen seconds, i.e. after four consecutive video chunks. Unlike the Vimeo and Dailymotion rate adaptive players in iPhones, Netflix is aggressive in providing the highest quality of the stream in Lumia825. In traffic traces, we noticed that the player begins with the lowest quality, and then switches to the maximum quality within the first few seconds of streaming. During this period, the player downloads 60s playback content. However, unlike the desktop player~\cite{Akhshabi:2012}, the mobile version requests different filenames for different qualities and specifies the byte range in the URL~\textsf{GET (abc).ismv/range/0-40140/}. In response, the server sends the chunks of the corresponding quality. The server also sends a \textsf{.bif} file which contains information about the frames, which is used by the player upon forward or backward seeking by the user. We also found that the Netflix server sends TCP parameters to the player.

\subsection{Summary}
Table~\ref{tab:mobile} summarizes our findings on the usage of different techniques in different mobile platforms with four video services. Figure \ref{fig:technique} illustrates how the client app behaviour leads to the choice of particular streaming technique for constant quality streaming. We sum up our main observations below:

\begin{figure}[t]
\centering
\includegraphics[width=0.4\textwidth]{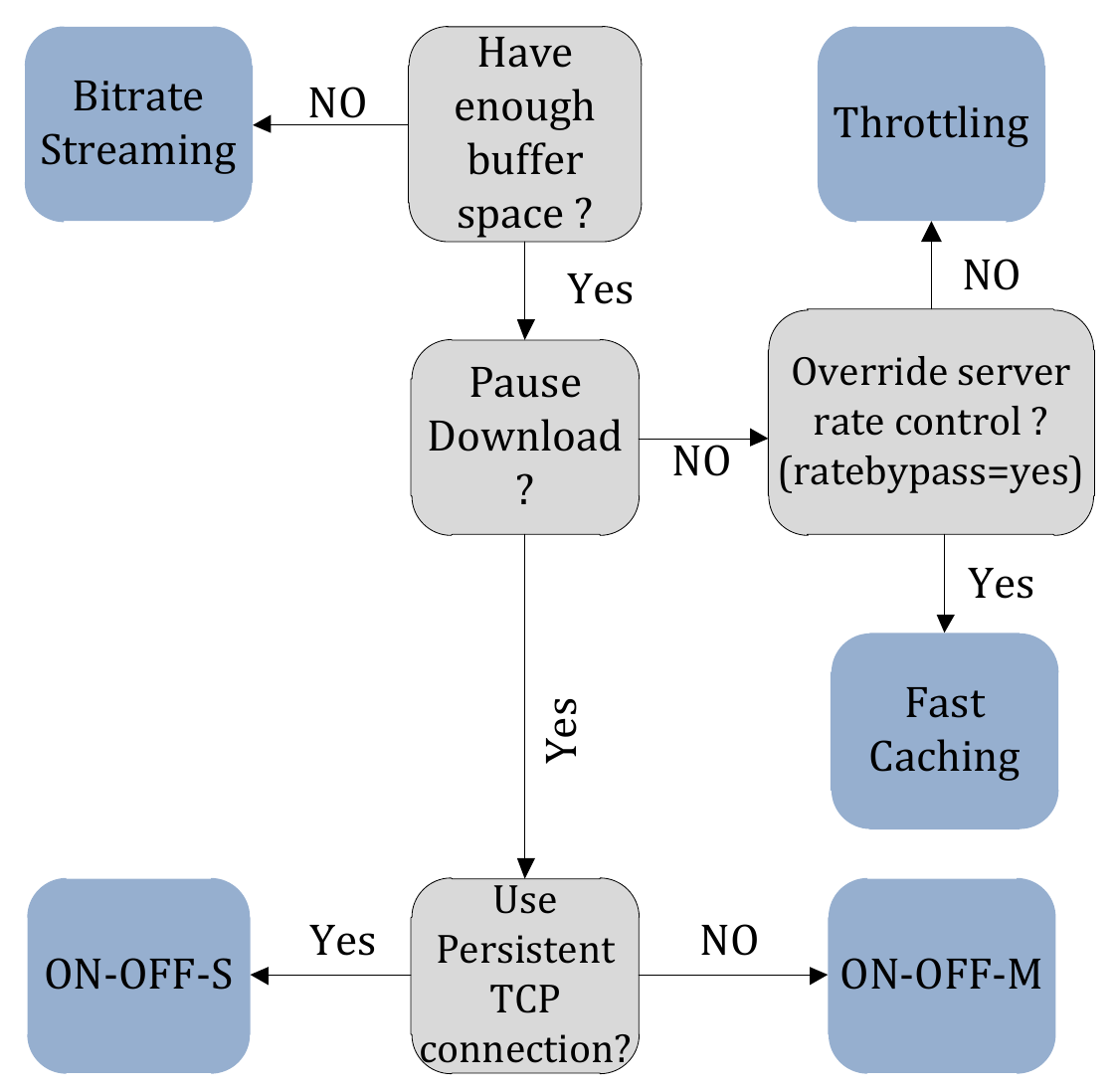}
\caption{The choice of a streaming technique by the client player for constant bit rate streaming.}
\label{fig:technique}
\end{figure}

\begin{figure*}[!t]
  \begin{center}
    \subfigure[Streaming Service]{\label{fig:service_join}
\includegraphics[width=0.24\linewidth]{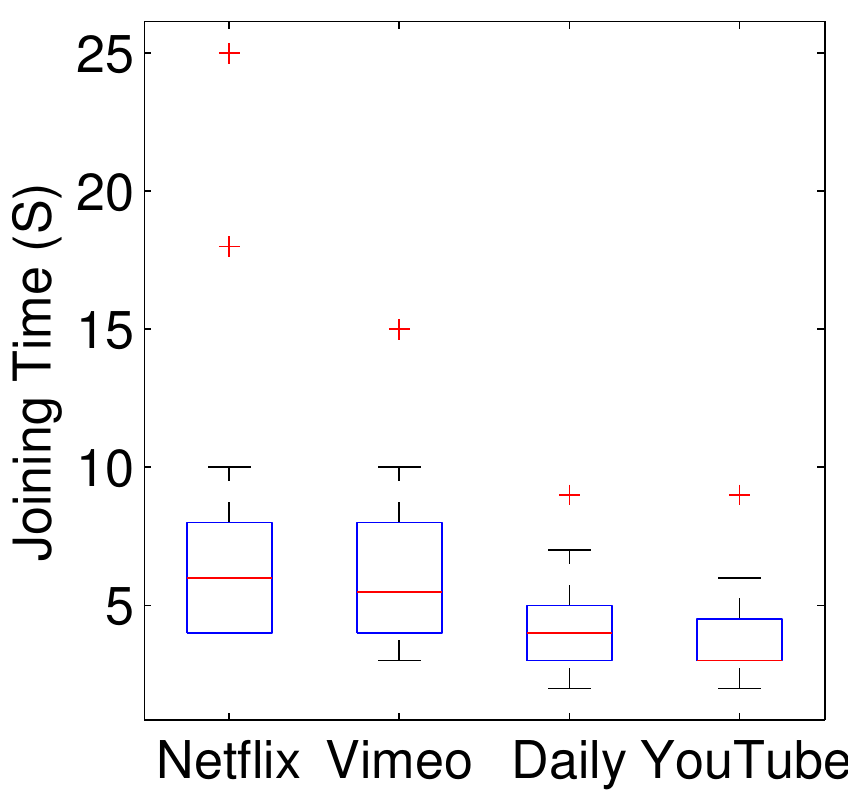}}
    \subfigure[Wi-Fi]{\label{fig:joining_time_wifi}
\includegraphics[width=0.24\linewidth]{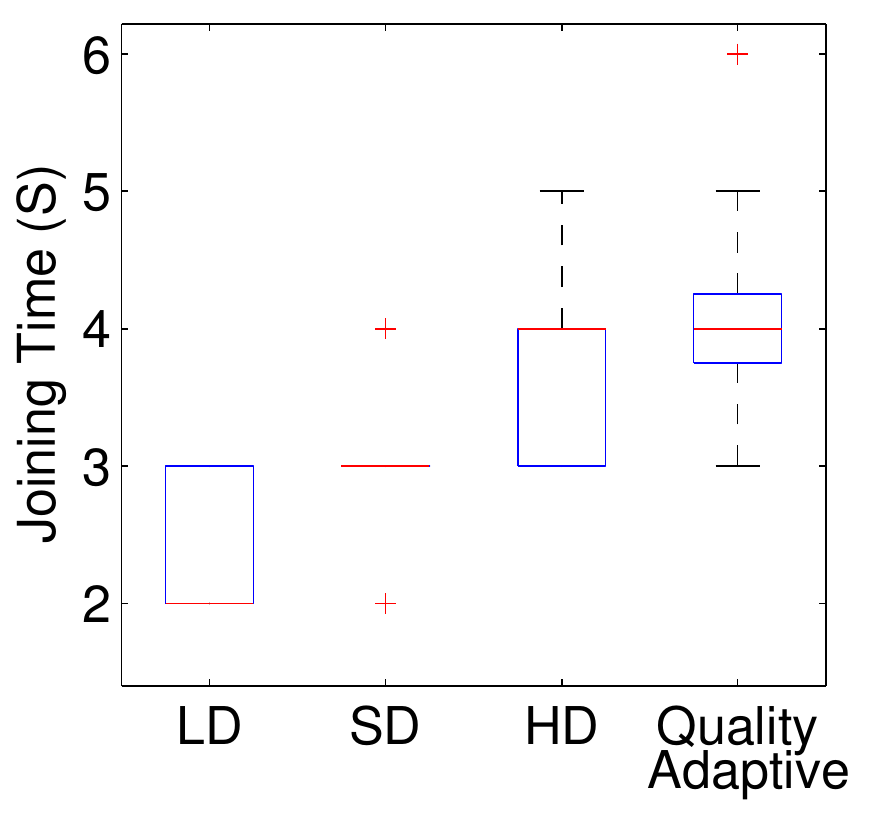}}
    \subfigure[HSPA]{\label{fig:joining_time_hspa}
\includegraphics[width=0.24\linewidth]{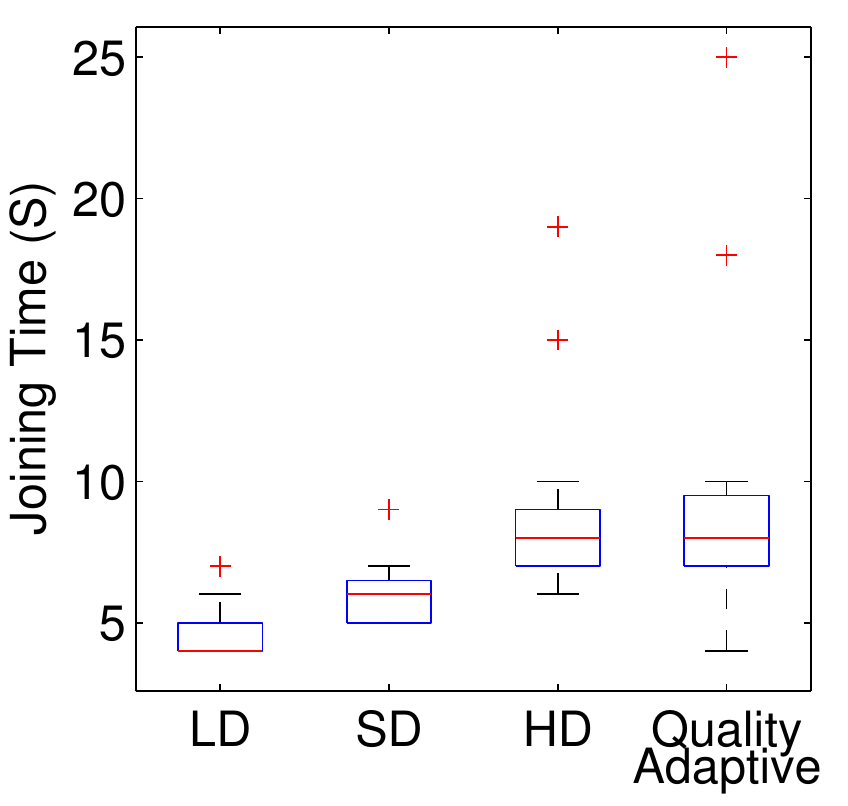}}
    \subfigure[LTE]{\label{fig:joining_time_lte}
\includegraphics[width=0.24\linewidth]{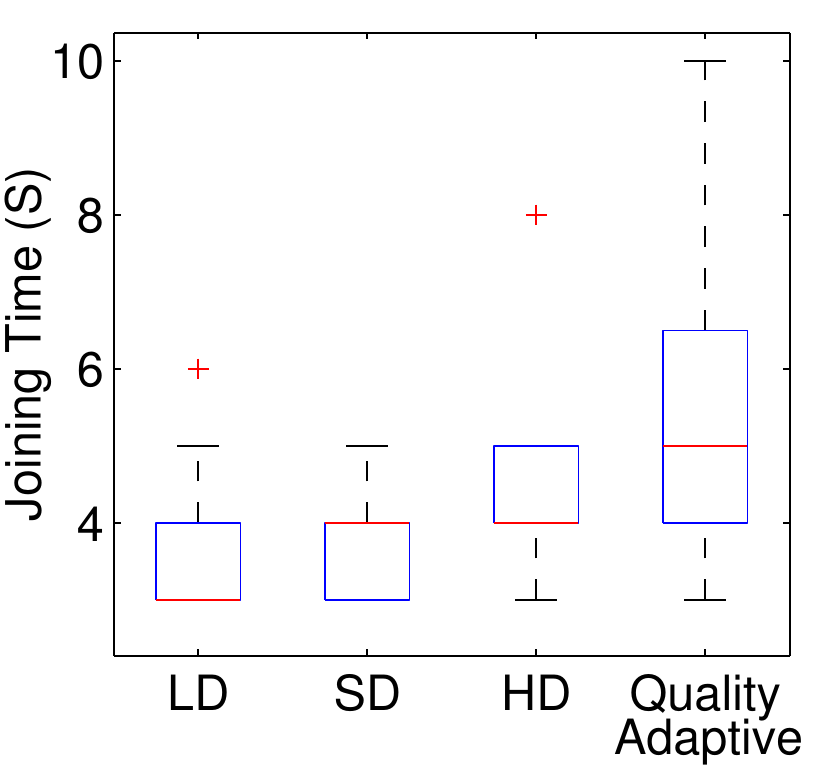}}
    \caption{Joining time observed for the video services and when streaming via wireless network interfaces.}
    \label{fig:joining_time}
 \end{center}
\end{figure*}

\begin{itemize}

\item Streaming servers use either throttling or fast caching to deliver constant bit rate video to mobile devices. The choice between these two is influenced by client player's request. For instance, YouTube Vimeo servers use both throttling and fast caching. The Dailymotion servers use throttling. Netflix servers use fast caching for constant bit rate streaming, and MSS or HSL for rate adaptive streaming. Some native mobile apps continuously pause and resume downloading leading to ON-OFF traffic patterns. Encoding rate streaming is the result of small playback buffer at the client buffer and fast caching streaming by the server.

\item For constant bit rate streaming, the relevance of a technique depends on the mobile platforms to some extent. Buffer adaptive streaming is commonly used by all the video streaming services in the Android platforms.  However, the only exception is Dailymotion. The reason is that the videos are small in size and the throttling rate is also small compared with YouTube and Vimeo. Therefore, the player does not get enough buffer filled to apply the adaptation. Fast caching is prevalent only in Windows-based devices. 

\item None of the video services apply rate adaptive streaming in Android mobile devices. The Netflix, Vimeo, and Dailymotion players use HLS in iPhones. In iOS devices, Netflix receives audio and video chunks in  separately streams. MSS is used only in Windows. This also reflects the influence of platforms on the choice of steaming techniques.

\item Although the streaming strategies can vary based on the quality of the video, platforms, and video services, we could not find any evidence that the strategies vary according to the wireless interface being used for streaming.


\begin{figure*}[!t]
  \begin{center}
    \subfigure[Encoding rate streaming and playback buffer status (in second).]{\label{fig:bitrate_buffer}\includegraphics[width=0.33\linewidth, height=0.26\textwidth]{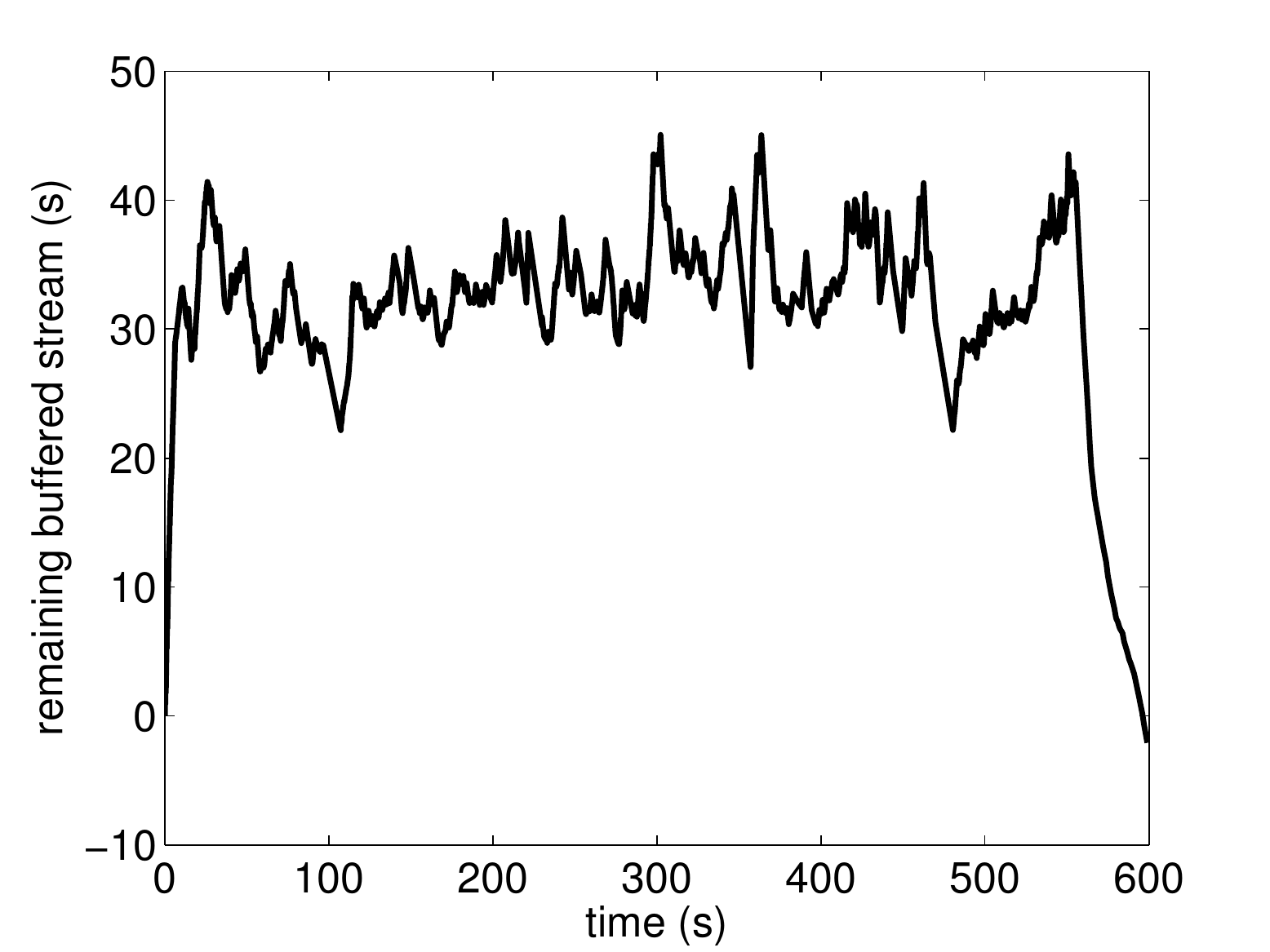}}
    \subfigure[Playback buffer status when using throttling, buffer adaptive and Fast Caching]{\label{fig:others_buffer}\includegraphics[width=0.3\linewidth, height=0.25\textwidth]{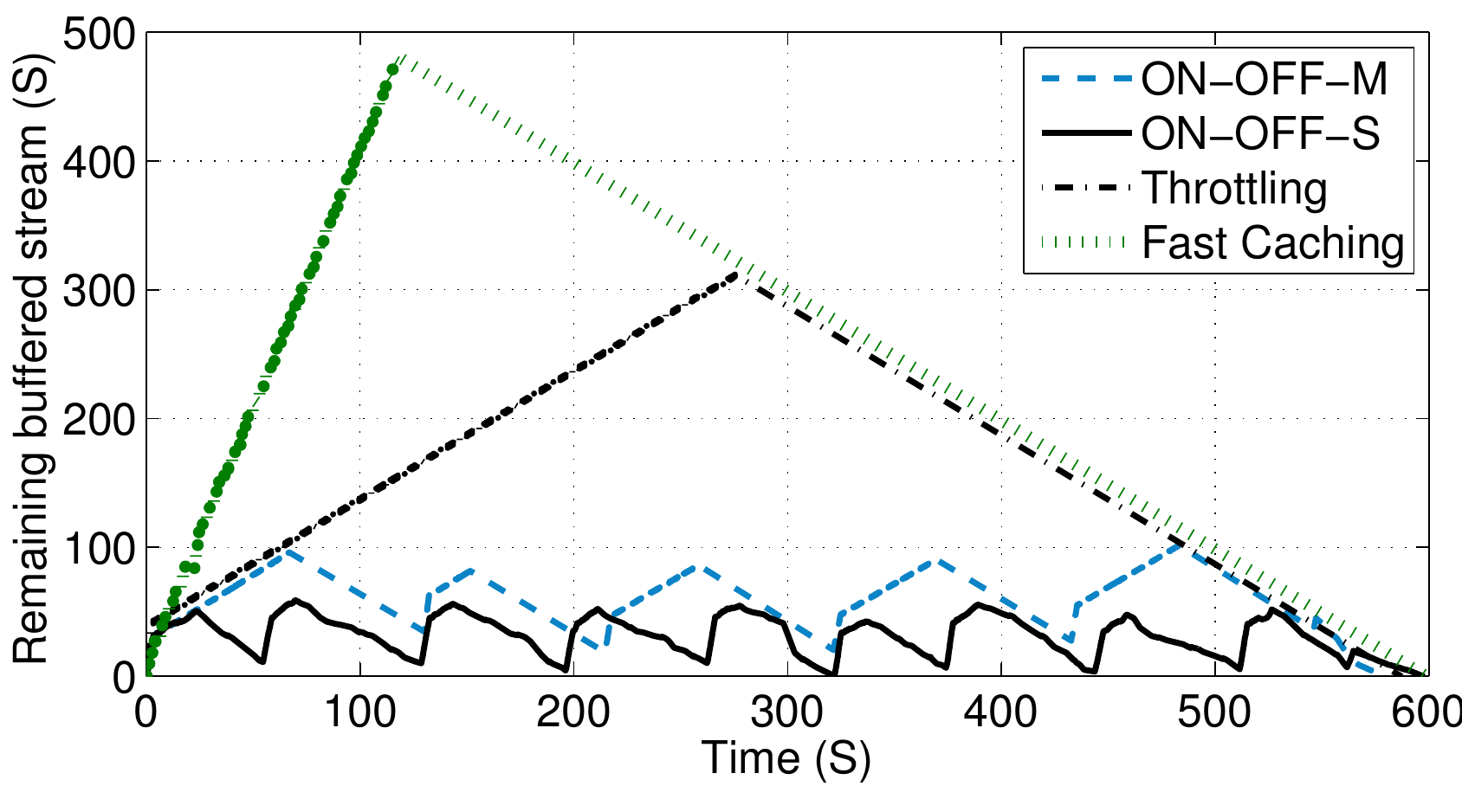}}
    \subfigure[Rate adaptive streaming and playback buffer status]{\label{fig:rateadap_buffer}\includegraphics[width=0.32\linewidth, height=0.25\textwidth]{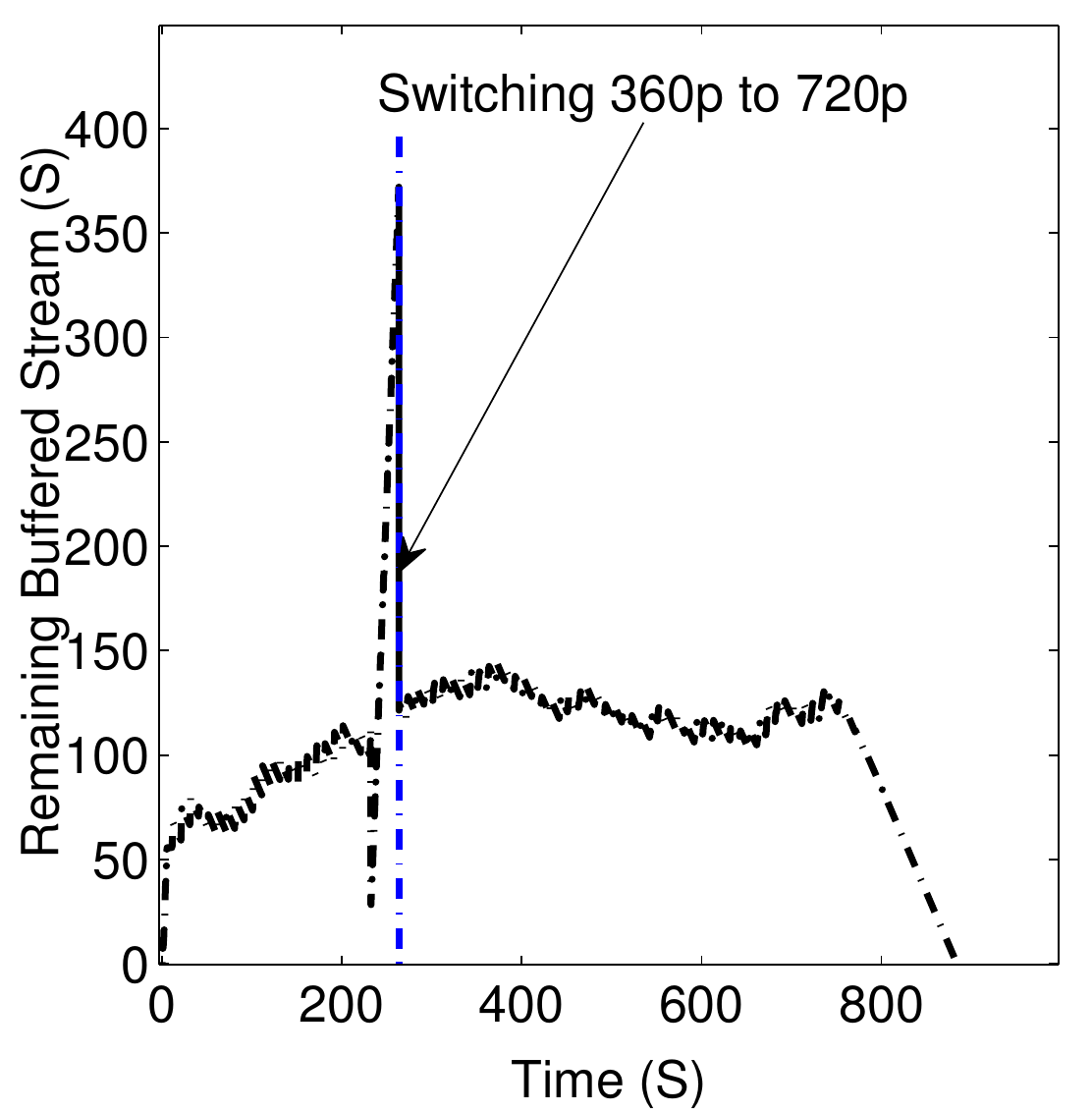}}
    \caption{Playback buffer status of the streaming clients during multimedia streaming sessions using different techniques.}
    \label{fig:playback_timeseries}
 \end{center}
\end{figure*}

\item The amount of data wasted by the YouTube player in iPhone4S is significant even when a user watches the complete video. Although this problem could be solved with a smarter player implementation, the YouTube player in the latest iOS version sends request with a throttle factor of 1.25. As a result, the playback buffer never becomes full, and consequently, there is no data waste. However, potential data waste is also possible when the whole video is downloaded and the user abandons watching earlier.
  

\end{itemize}

\section{Streaming Techniques and Quality of Experience}
\label{qoe_streaming}
The key metrics that characterize the QoE perceived by a user while streaming video are the initial playback delay which is called joining time, and the occurrence and frequency of playback pause events experienced~\cite{Balachandran:2012,Balachandran:2013}. As we discussed earlier that playback pause events are the results of bandwidth variation due to various network conditions. In this section, we take a look at the joining time and the performance of different strategies in providing smooth playback during short/long term bandwidth changes.





Figure~\ref{fig:joining_time} shows the joining time experienced by the players according the video service and the WNI. Although, all the streaming services use fast start at the beginning of streaming, it is shown in Figure~\ref{fig:service_join} that the YouTube players take less time than the other players. On the other hand, the other services have longer joining time. The reason is that YouTube caching servers are extensively spread around the globe. Therefore, the content is served from the CDN that is very close to the user. We validated this by measuring the roundtrip time from the captured traces. Our observation is similar to \cite{Balachandran:2012}, in which the authors also proposed to serve content from the nearby CDN to improve the playback experience. However, in the case of Vimeo and Netflix two other facts also contribute in higher joining time. The Vimeo player always receives the HD quality video and the Netflix player always decides the maximum quality at the beginning of streaming, which take more time than the players of other services. 

We explained earlier that quality of the stream affects the initial start-up time. The boxplots in Figure~\ref{fig:joining_time_wifi},~\ref{fig:joining_time_hspa},~\ref{fig:joining_time_lte} illustrate the similar findings. There are two observations. First, streaming via Wi-Fi experiences less joining time than streaming via HSPA and LTE. The joining time is the largest when HSPA is used. We investigated and found that the wireless latency plays the role when streaming via HSPA and LTE. This is because, at the beginning of a streaming session, the HSPA interface transitions from IDLE/CELL\_PCH to CELL\_DCH state and the LTE interface switches from IDLE to the CONNECTED state. The transition latencies for LTE and HSPA are 120ms and 2.0s respectively. In the case of Wi-Fi, the transition latency from sleep to active state transition is few milliseconds which is negligible. The other observation is that the rate adaptive players experience more delay in the joining. This observation is biased because of the Netflix's rate switching strategy.

Next, we looked at the prefetching behavior of the players by studying how much content they maintain in the playback buffer throughout a streaming session. This analysis requires the time series of content consumption and arrival. The arrival time series is computed by extracting timestamps and playload sizes of received packets from the traffic traces considering the joining time. Although there are findings that YouTube-like video services stream constant bit rate content~\cite{Chengx:2013}, we found that the video services use variable bit rate encoding for streaming HD videos. Hence, we replayed each video using a VLC player and extracted the instantaneous encoding rate of the content from VLC's web interface module using a shell script. Finally, we compute the amount of buffered content as a function of time by taking the difference of the cumulative sums of the arrival and consumption time series.

Figure~\ref{fig:playback_timeseries} shows the playback buffer status during the streaming sessions using different streaming techniques. Using encoding rate streaming, a player always keeps 30-40s equivalent content in the playback buffer. Hence, even if the player receives content at the negligible rate after the fast start phase, the player can provide playback for that 30-40s period. Throttling and fast caching continuously accumulate more content into the buffer and therefore are more robust also towards longer periods of low available bandwidth. From~\ref{fig:others_buffer}, we can see that when the playback is at 50s, the player already has content for next 50s using throttling. In case of fast caching, the player has 200 s worth of content in the buffer. When using the ON-OFF strategies, the buffer is periodically filled up and drained in between. ON-OFF-M begins refilling the buffer 40s earlier. A surprising result is that ON-OFF-S (in Android 2.3.6) nearly dries the buffer before new content is prefetched. Therefore, the possibility of playback starvation increases, when streaming via HSPA. The rate adaptive players maintain 60-100s playback buffer, and at the same time they can select to a lower quality (see Figure~\ref{fig:rateadap_buffer}). Nevertheless, the streaming strategies provide the best effort in guarding short term and long term bandwidth fluctuations.

\section{Streaming Services and Power Consumption}
\label{five}

We also measured the total current consumed by the smartphones during the streaming sessions. We separated the total current drawn into the average video playback and wireless interface current consumption. The playback current includes decoding and display current. We can identify this current draw at the end of the power trace of each streaming session when the content has been fully delivered but playback still continues, since some of content is always buffered at the end regardless of streaming technique used. During this time, the WNIs are in the lowest power consuming states according to their own power savings protocols. We computed the average wireless communication current,
which we refer to as streaming current, by subtracting the average
playback current from the total current. The results presented in this section are the average of repeated measurements.

\begin{figure}[!t]
  \begin{center}
    \subfigure[Avg. playback current draw when streaming $240-1080p$ YouTube videos to the app and browser in Galaxy S3.]{\label{fig:power_youtube_all_galaxy}
\includegraphics[width=0.9\linewidth, height=0.24\textwidth]{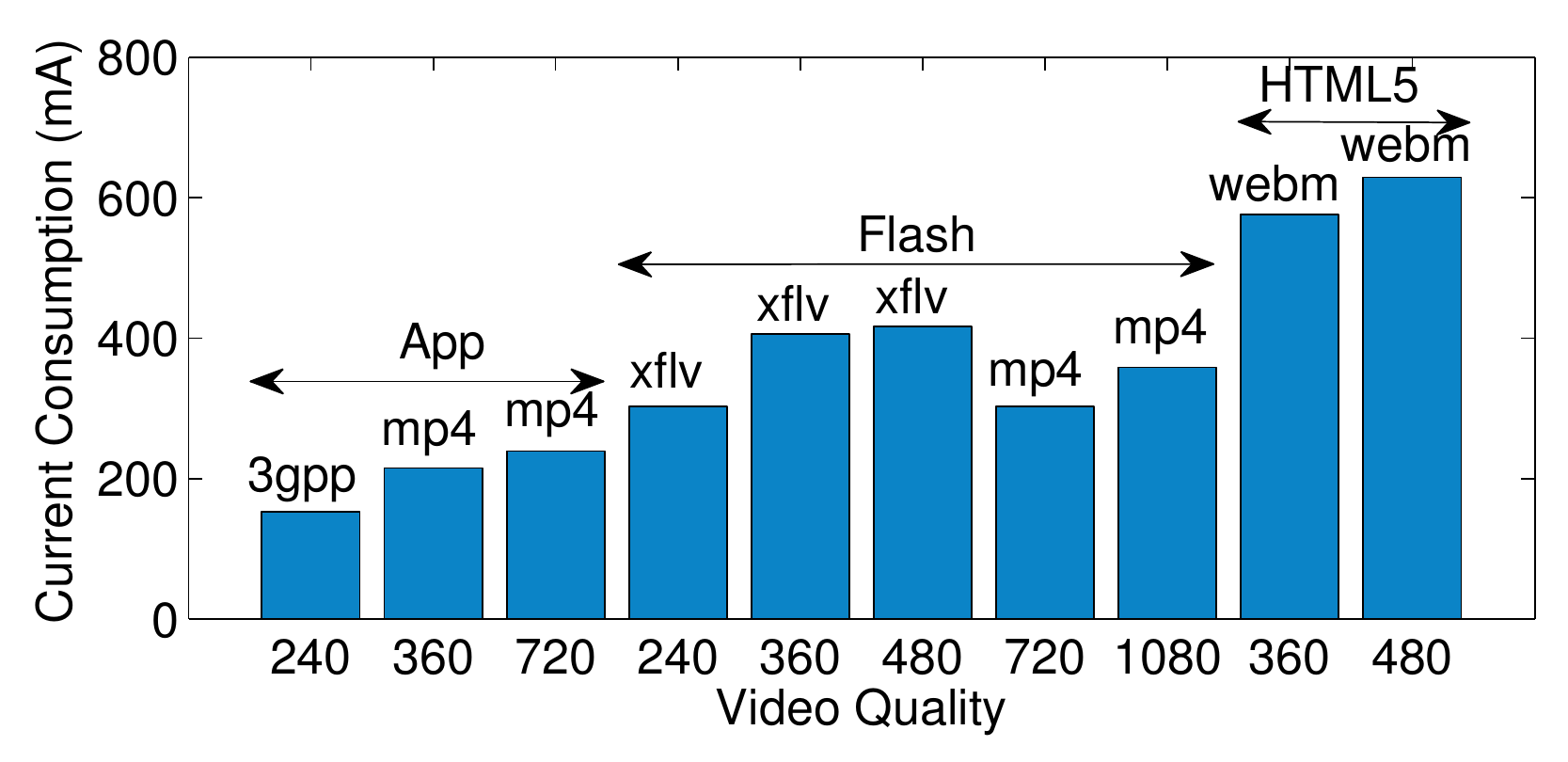}}
        \subfigure[Amount of CPU used by different video players in Galaxy S3 while playing different quality videos of different containers.]{\label{fig:cpu_usage}
\includegraphics[width=0.9\linewidth, height=0.23\textwidth]{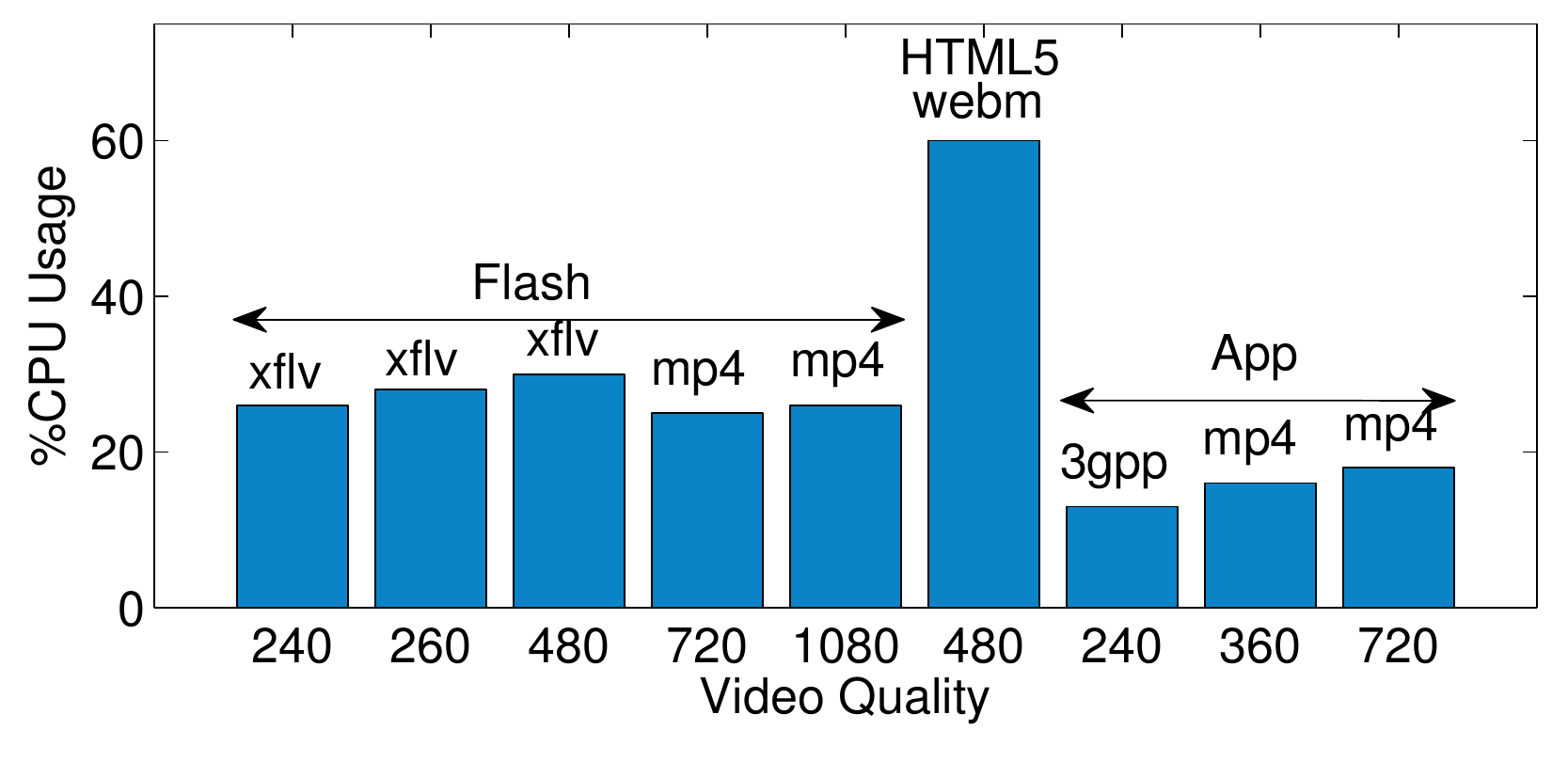}}
    \subfigure[Avg. playback current consumption while playing different quality videos of different containers.]{\label{fig:container_all}\includegraphics[width=0.9\linewidth, height=0.23\textwidth]{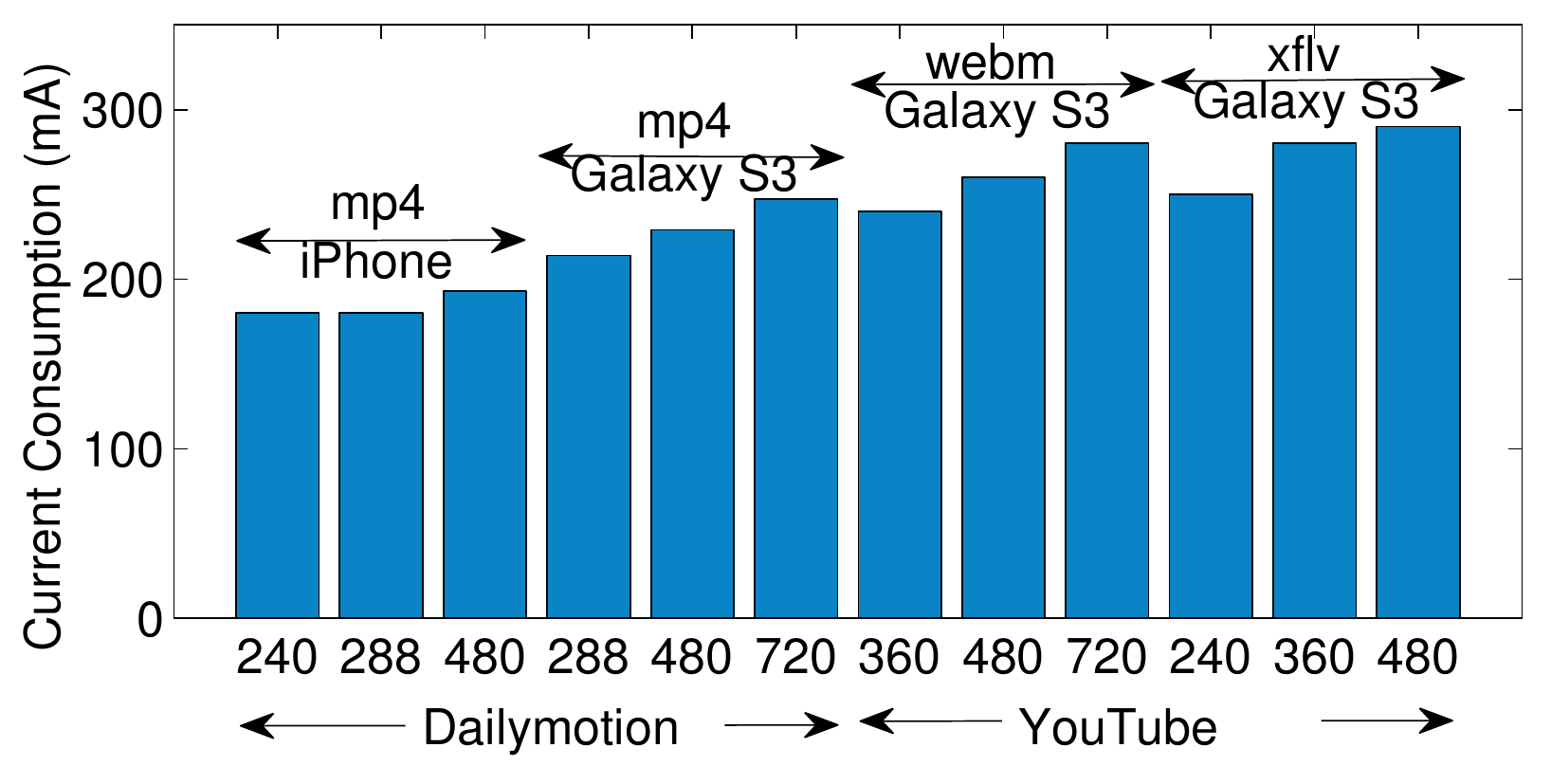}}
    \caption{Playback current consumption of Galaxy S3 and CPU usage with different qualities, players and containers.}
    \label{fig:quality_container}
 \end{center}
\end{figure}

\subsection{Playback Power Consumption}
\label{five_one}

\subsubsection{Video Quality} In Figure~\ref{fig:power_youtube_all_galaxy}, we can see that playback current draw of Galaxy S3 increases as the quality of YouTube video increases as long as the same container is used. We also observed similar pattern for watching Dailymotion videos in iPhone4S and Galaxy S3. It is logical that high quality videos have more information to present than low quality videos and, therefore, more current is drawn. However, in some cases even doubling the resolution adds a relatively small increment to the average playback current.

\begin{figure}[!t]
  \begin{center}
    \subfigure[Wi-Fi and HSPA]{\label{fig:wifi_hspapower}\includegraphics[width=0.7\linewidth, height=0.25\textwidth]{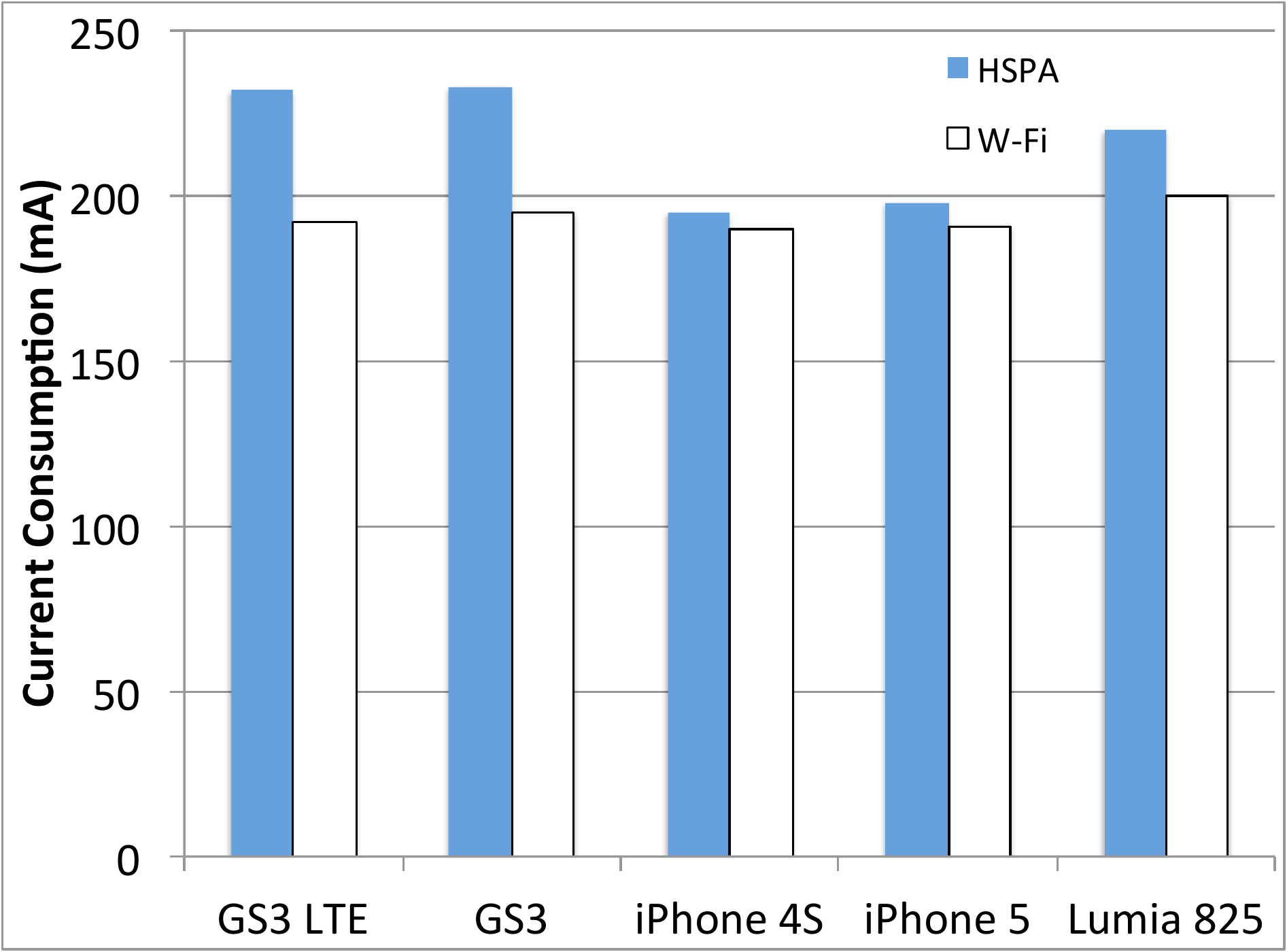}}\\
    \subfigure[LTE]{\label{fig:lte_rrc_config}\includegraphics[width=0.7\linewidth, height=0.25\textwidth]{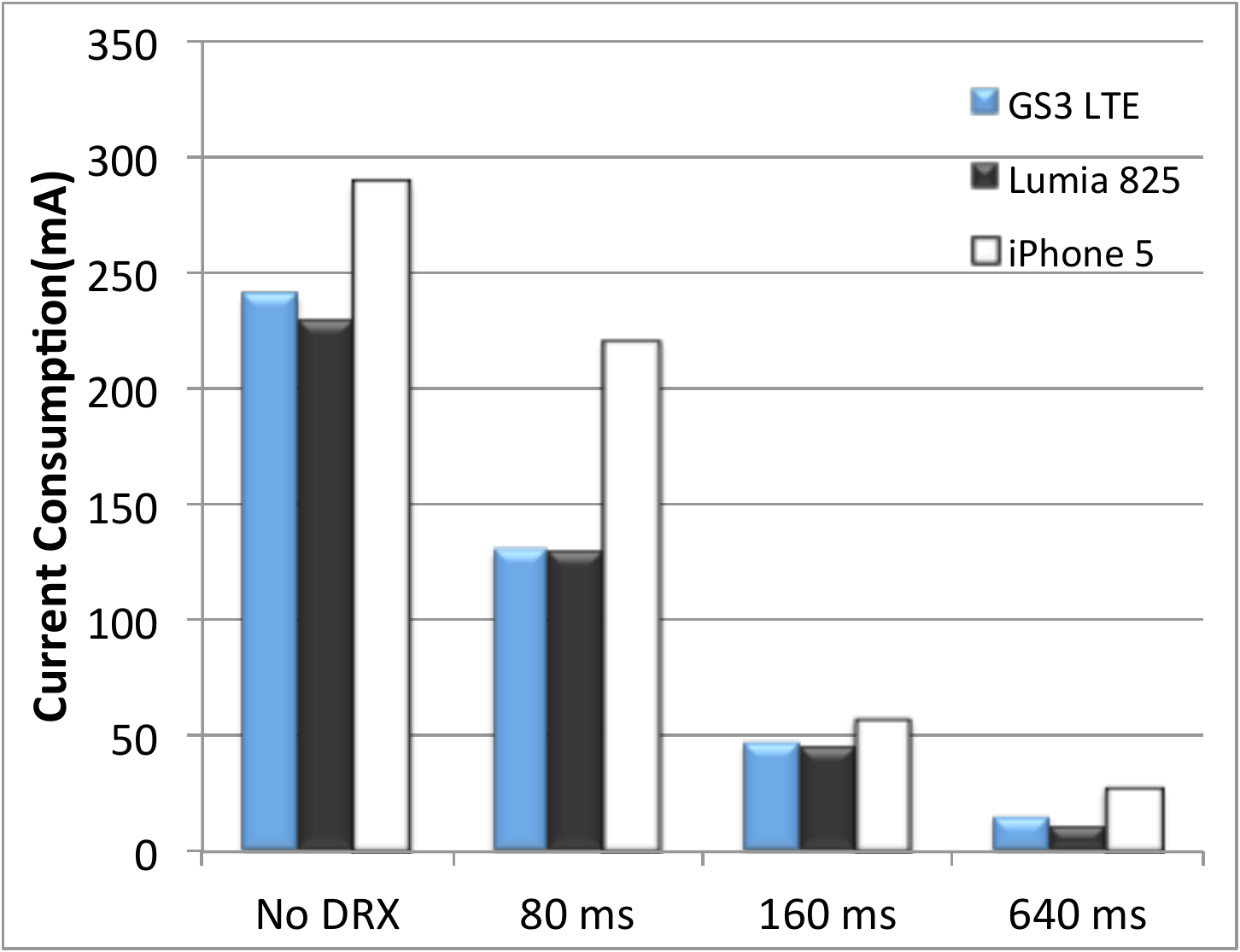}}
    \caption{Current consumption of wireless network interfaces in smartphones.}
    \label{fig:interfacepower}
 \end{center}
\end{figure}

\subsubsection{Video Player} 
For playing YouTube LD, SD and HD videos, the browser loads a Flash player. Flash has support for different kind of codecs and containers, such as X-FLV, MP4 and H.264. The browser loads HTML5 player to play WebM videos. Figure~\ref{fig:power_youtube_all_galaxy} compares the energy consumption when using different players for streaming. It is noticeable that the native YouTube application consumes the least amount of energy. In contrast, browser-based players can draw even more than the double current compared with the app when playing the same video. We discovered that during playback the Flash player does not leverage any native system support to decode the video but consumes a significant amount of more CPU than the native application (see Figure~\ref{fig:cpu_usage}). Although the HTML5 player takes native system support, it consumes 60\% of CPU even during the playback of a $480p$ video. It seems that HTML5 player is required to go through further optimization to be used in mobile platforms.

\subsubsection{Video Container}

We already showed how the videos of different quality and different players affect the energy consumption of smartphones. In Figure~\ref{fig:power_youtube_all_galaxy}, we can see that playback of a $240p$ 3GPP video requires less energy than that of an X-FLV video of the same quality. It is also illustrated that the same $240p$ X-FLV requires more current than a $720p$ MP4 video. Although from Figure~\ref{fig:power_youtube_all_galaxy} we can infer that 3GPP is the least and WebM is the most energy consuming containers, it is difficult to isolate the effect of the corresponding video containers since some videos can be played only using browsers. Besides, the energy consumption of the browser-based players are very high. Therefore, we downloaded some YouTube videos of X-FLV and WebM formats and then measured energy consumption during playback. The results are shown in Figure~\ref{fig:container_all}. This figure also illustrates that playback energy consumption does not change significantly when the quality of video changes with the same container category.

\subsection{Device Variation}

Before discussing the impact of different streaming strategies on the streaming power consumption, we investigate the power consumption of individual WNI in smartphones. In Section~\ref{sec:powersave}, we described the standard power saving mechanisms applied by different WNIs. We also discussed that there are a number of states and a mobile device consumes different amount of energy in different states. Consequently, we explore what kind of power saving mechanism are applied by our target smartphones and the variation among them in consuming energy.

In Figure~\ref{fig:wifi_hspapower}, we can see that the Wi-Fi interfaces in iOS phones consume lowest energy. Android devices consume more current when the Wi-Fi interface is active, whereas the Wi-Fi interface in Lumia825 consumes the maximum energy.  However, all of them use PSM adaptive. iOS devices use an aggressive idle period of 50ms. The other devices use 200ms. The power consumption during this idle state is half of the active state power consumption. Figure~\ref{fig:wifi_hspapower} shows the power consumption of HSPA interface during data transfer in CELL\_DCH state. In this case of also iOS devices consume the lowest energy when the HSPA interface is active. Lumia825 is the second least. On the other hand Android devices consume the maximum energy. However, all the devices use Fast Dormancy with an inactivity timer of 5s, except iPhone5 which uses an inactivity timer of 8s.

We measured power consumption of the LTE interface with four different network configurations; DRX is disabled, DRX is enabled with a short DRX cycle (80ms), with DRX cycles of 160ms and 640ms respectively. From the results presented in Figure~\ref{fig:lte_rrc_config}, we find that the smartphones consume the maximum energy when DRX is not enabled in the network. If DRX is enabled in the network, the smartphones consume less power. This is because the devices periodically wake up to check data activity according to the DRX cycles in the connected state. This Figure also depicts that Lumia825 consumes the lowest current when LTE is active.

\begin{figure}[!t]
  \begin{center}
    \subfigure[DRX cycle length 80 ms.]{\label{fig:short_drx}\includegraphics[width=0.49\linewidth, height=0.25\textwidth]{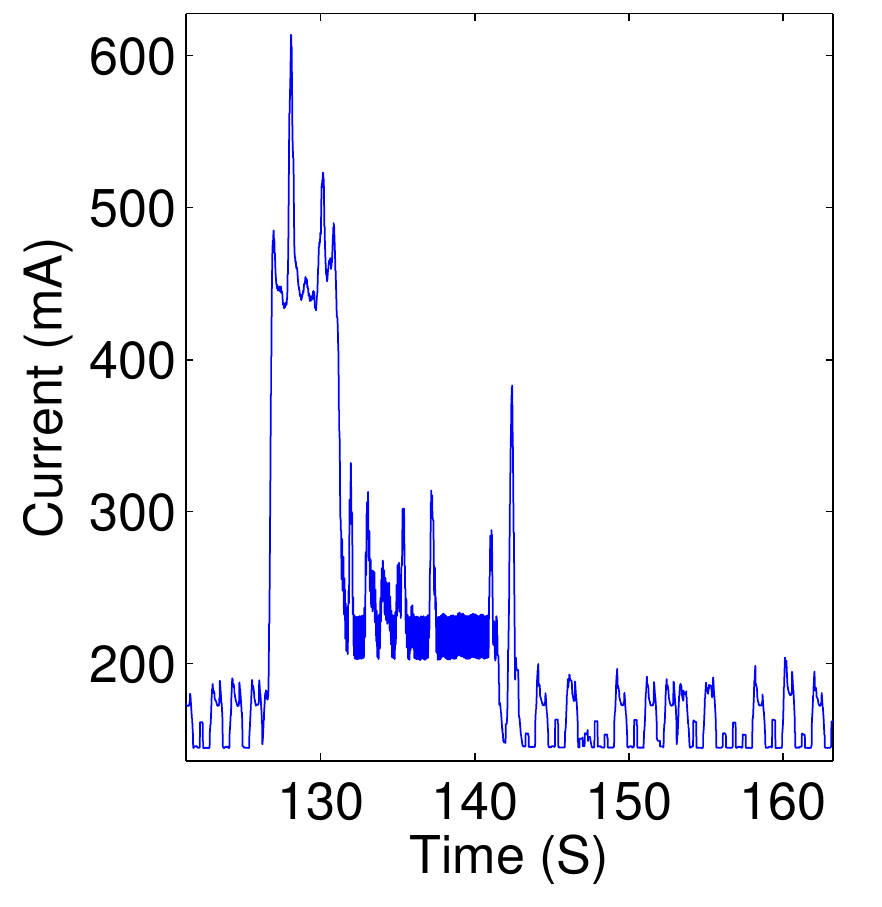}}
    \subfigure[DRX cycle length 640 ms.]{\label{fig:long_drx}\includegraphics[width=0.49\linewidth, height=0.26\textwidth]{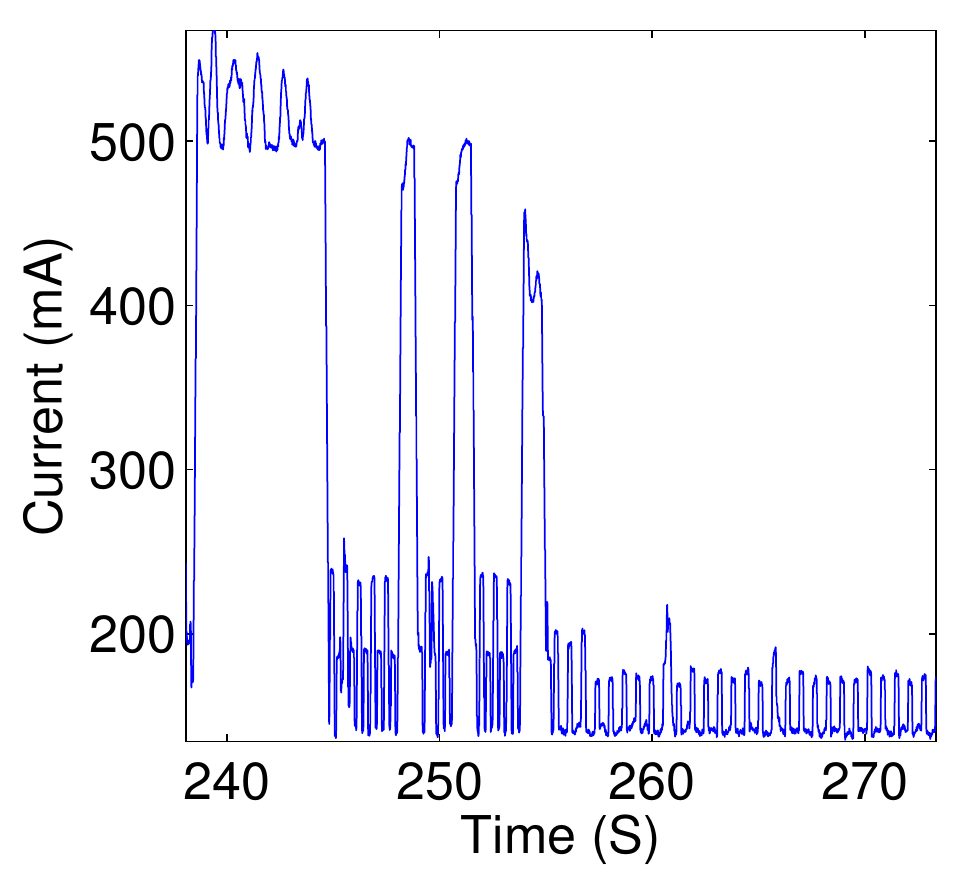}}
    \caption{Current consumption of GS3 LTE with different DRX cycles.}
    \label{fig:ltepower}
 \end{center}
\end{figure}

\begin{figure*}[!t]
\centering
\includegraphics[scale=0.35]{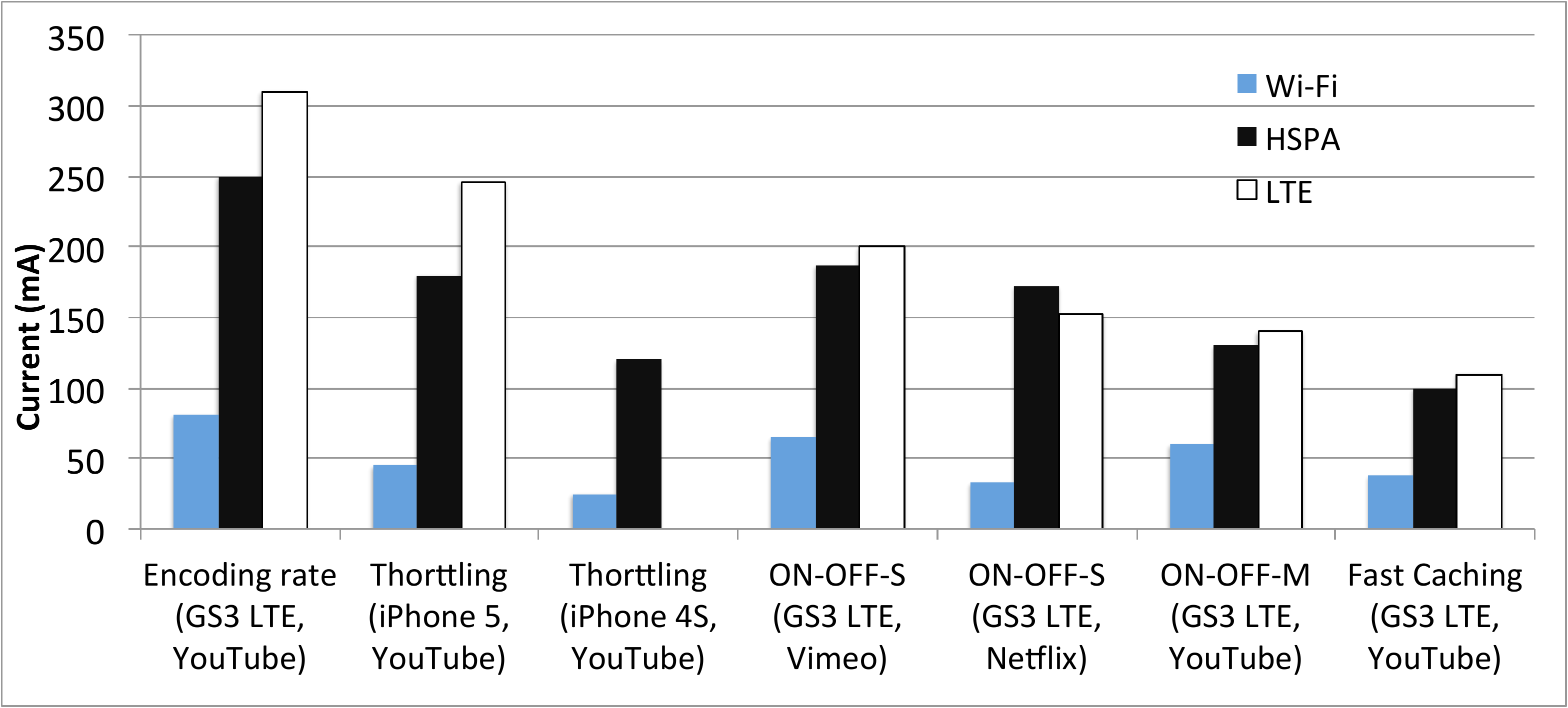}
\caption{Avg. streaming current consumption of smartphones when streaming a 600 s long constant bit rate video using the streaming strategies.}
\label{fig:wireless_current}
\end{figure*}

Figure~\ref{fig:lte_rrc_config} also depicts that iPhone5 is the most and Lumia825 is the least energy consuming device when DRX is enabled. From power traces we identified that even though the DRX$_{on}$ was configured to 10ms, iPhone5 spends 60ms. On the other hand, Lumia825 and GS3 LTE spend 30 and 45ms respectively in the on period of the DRX cycle. From Figure~\ref{fig:lte_rrc_config}, we can also see that the devices consume more current when the cycle lengths are shorter. For instance, when DRX cycle is of 80ms, GS3 LTE and Lumia825 consume around 120mA current. If the cycle length is increased to 640ms, the power consumption is decreased by a factor of three approximately. The first reason is that when short DRX cycles are in action, a mobile device will spend more time in the on period of the cycles as there will be more cycles when the RRC inactivity timer is active. Second, the LTE chipset is not optimized yet to operate on such small cycles. They cannot efficiently shutdown the power consumption during the DRX sleep phase. Figure~\ref{fig:short_drx} shows that current consumption of GS3 LTE is stable at $\approx$220mA from 132 to 142s even though the DRX is active. Current consumption during short DRX cycles does not scale down like when DRX cycle is of 640ms (from 245 to 255s in Figure~\ref{fig:long_drx}). This pattern is also consistent with iPhone5 and Lumia825 (Figure~\ref{fig:lte_rrc_config}). 


However, power consumption of these interfaces can vary according to the downloading rate. The deviation can be $\pm 50$mA.

\subsection{Impact of Streaming Techniques}
\label{five_three}

In the previous section, we showed the basic power consumption characteristics of different WNI. In this section, we discuss the effect of streaming techniques on the energy consumption in smartphones. Since all the techniques are not available in a single platform, it is difficult to compare the energy efficiency of the techniques. Therefore, we compare only the current consumed by the wireless interfaces of the smartphones and exclude the playback current in order to provide a comparison ground. In the case of LTE, the DRX was enabled in the network and we used a single DRX profile with DRX cycle of 80ms, as this profile is used by the network operators in Finland. We compare them in Figure~\ref{fig:wireless_current}. 

\subsubsection{Encoding Rate Streaming} In this case, the content is delivered continuously throughout the entire streaming session and the wireless interface is active all the time. For example, downloading a 6 minute video would require approximately six minutes. As a consequence, the average streaming current drawn by Galaxy S3 LTE is very high for the YouTube videos. Figure~\ref{fig:wireless_current} also shows that Galaxy S3 LTE (GS3 LTE) consumes around 77mA for Wi-Fi, 200mA and 310mA for HSPA and LTE respectively (HD video using browser). The high current consumption of HSPA/LTE is expected, since these interfaces are constantly in the highest power consuming state. However, power consumption over Wi-Fi is low with respect to the usage of the interface. This is because, the Android devices use DVFS when streaming via Wi-Fi.


\subsubsection{Throttling} 

In Section~\ref{four_two}, we discussed that in case of throttling, the throttle factor defines the amount of time is used to deliver the content to the client players. The higher is the throttle factor, the lower is the time required at the client to download the content. Therefore, this factor also determines the amount of time the wireless radio will be powered on and hence it also determines power consumption at smartphones. Energy consumption for two throttled sessions is presented in Figure~\ref{fig:wireless_current}. In the first case, the server uses the throttle factor 1.25 for iPhone5. The second session is for iPhone4S, where the factor is 2. iPhone5 consumes more current than iPhone4S for streaming via Wi-Fi and 3G. The obvious reason is that iPhone4S downloads at a faster rate. And both smartphones consume less current than the GS3 LTE which downloads video at the encoding rate. Therefore, throttling delivers energy savings over encoding rate streaming as interface usage time is reduced.

\subsubsection{Buffer Adaptive Streaming} 
\label{on-off-energy}

Figure~\ref{fig:wireless_current} shows that GS3 LTE consumes more current in streaming a Vimeo video than the Netflix video via any WNI. This is because of the player behavior in resetting TCP persist timer. We described in Section~\ref{on-off-s} that the Vimeo player resets TCP persist timer after every 16 seconds. Therefore, the maximum interval between TCP control packets from Vimeo can be 5s. On the other hand, the Netflix player rests after every 30s and the maximum interval between TCP control packets from Netflix is 10s. Therefore, the interfaces can spend more time in low power consuming states when streaming from Netflix than streaming from Vimeo. However, the average streaming current consumption is less than the encoding rate streaming.

Figure~\ref{fig:wireless_current} also includes a case where GS3 LTE receives content from YouTube in multiple TCP connections. Since the duration of such an OFF period is 60s, the wireless interfaces can be in sleep or the lowest power consuming states for very long time. As a result, GS3 LTE consumes roughly 50\% less energy when using ON-OFF-M than the encoding rate. However, it can be seen that ON-OFF-M does not outperform throttling (iPhone4S) in current consumption as the player receives content at the same throttled rate in each TCP connection.



\begin{figure}[!t]
\centering
\includegraphics[width=0.7\linewidth]{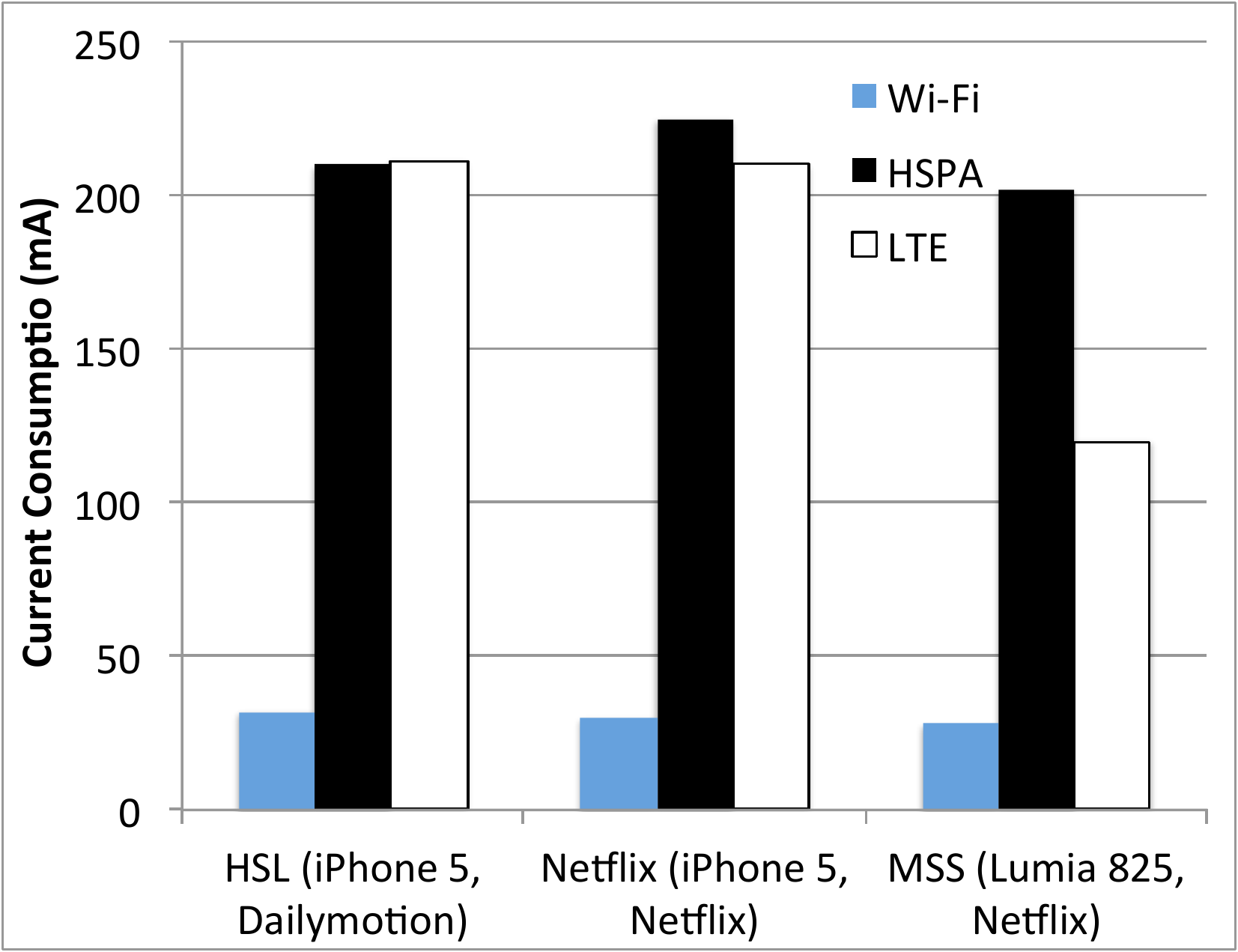}
\caption{Avg. streaming current consumption of smartphones for rate adaptive streaming techniques, HTTP Live Streaming, Microsoft Smooth Streaming and Netflix's own adaptive mechanism in iPhone5.}
\label{fig:wireless_current_two}
\end{figure}

\subsubsection{Fast Caching}
Fast caching is used to download content at the client with as high throughput as possible. As a result the wireless interface is maximally utilized for as little time as possible. Figure~\ref{fig:wireless_current} shows that GS3 LTE consumes  the least current, if the YouTube player downloads the whole video using Fast Caching. 

\subsubsection{Rate Adaptive Streaming}
Similar to the ON-OFF-M mechanism, the quality or rate adaptive players also receive content in chunks over a single or multiple TCP connections. The duration of a chunk varies from a minimum four seconds to maximum ten seconds depending on the service. Figure~\ref{fig:wireless_current_two} shows the current consumption of the WNIs when streaming Netflix and Dailymotion videos in iPhone5 and Lumia825. In both devices, power consumption of the Wi-Fi interface is about 30mA. The players in iPhone receive content in 10s chunks. Therefore, the HSPA interface avails the lower states rarely as the FD timer is 8s and consequently current consumption is high. The LTE interface also consumes significant current even though the DRX was enabled. This is because, the LTE interface in the iPhone5 takes long time in the ON period of the DRX cycle. iPhone5 consumes more current when streaming Netflix than the Dailymotion via cellular networks. The reason is that the Netflix player downloads audio and video chunks separately and their downloading was not synchronized. Compared with iPhone5, Lumia825 consumes less current when the Netflix player streams via LTE as the interface spends lesser time in the ON state of the DRX cycles when DRX is active.

\subsection{summary}

From Section~\ref{five_one}, we learned that native apps are the most energy efficient. Since, HTML5 is an important technology at this moment, optimizing the HTML5-based player implementations would be an important future work. We also noticed that video container/codec also has significant impact on the energy consumption (3GPP seems more efficient than MP4), while video quality has a small impact. Therefore, the focus should be choosing an optimal codec or container.

Concerning the current consumption of wireless network interfaces, Wi-Fi is the the most energy efficient interface. When using LTE, the smartphones are not optimized yet for 80-160ms DRX cycles. Therefore, the network operators should use longer DRX cycles in the network to improve the battery life time of smartphones. The main lesson concerning the different streaming techniques is that encoding rate streaming causes clearly the largest amount of energy consumption. Fast caching is the most energy efficient technique. An effective ON-OFF-M technique should deliver content without any rate control. Although the rate adaptive techniques are similar to ON-OFF-M, higher chunk size and synchronization between audio/video chunks would reduce energy consumption significantly.


\section{QoE and Energy Consumption Tradeoffs}
\label{tradeoff}

In Section~\ref{qoe_streaming}, we found that most of the video services use optimized methods so that streaming quality does not deteriorate user experience by enabling the players in providing uninterrupted playback as long as possible. From this perspective, fast caching and throttling are the most efficient techniques. However, if the user does not watch the whole video, the downloaded data is wasted. Furthermore, using the cellular access to download unnecessarily content is problematic for users having small quota in their data plan and for the network resources. For example, Finamore et al. analyzed YouTube traffic to desktop computers and iOS devices accessed via Wi-Fi and discovered that 60\% of videos were watched for less than 20\% of their duration~\cite{Finamore:2011}. Therefore, ON-OFF mechanisms are attractive considering the unnecessary content download.
\begin{figure}[!t]
\centering
\includegraphics[scale=0.5]{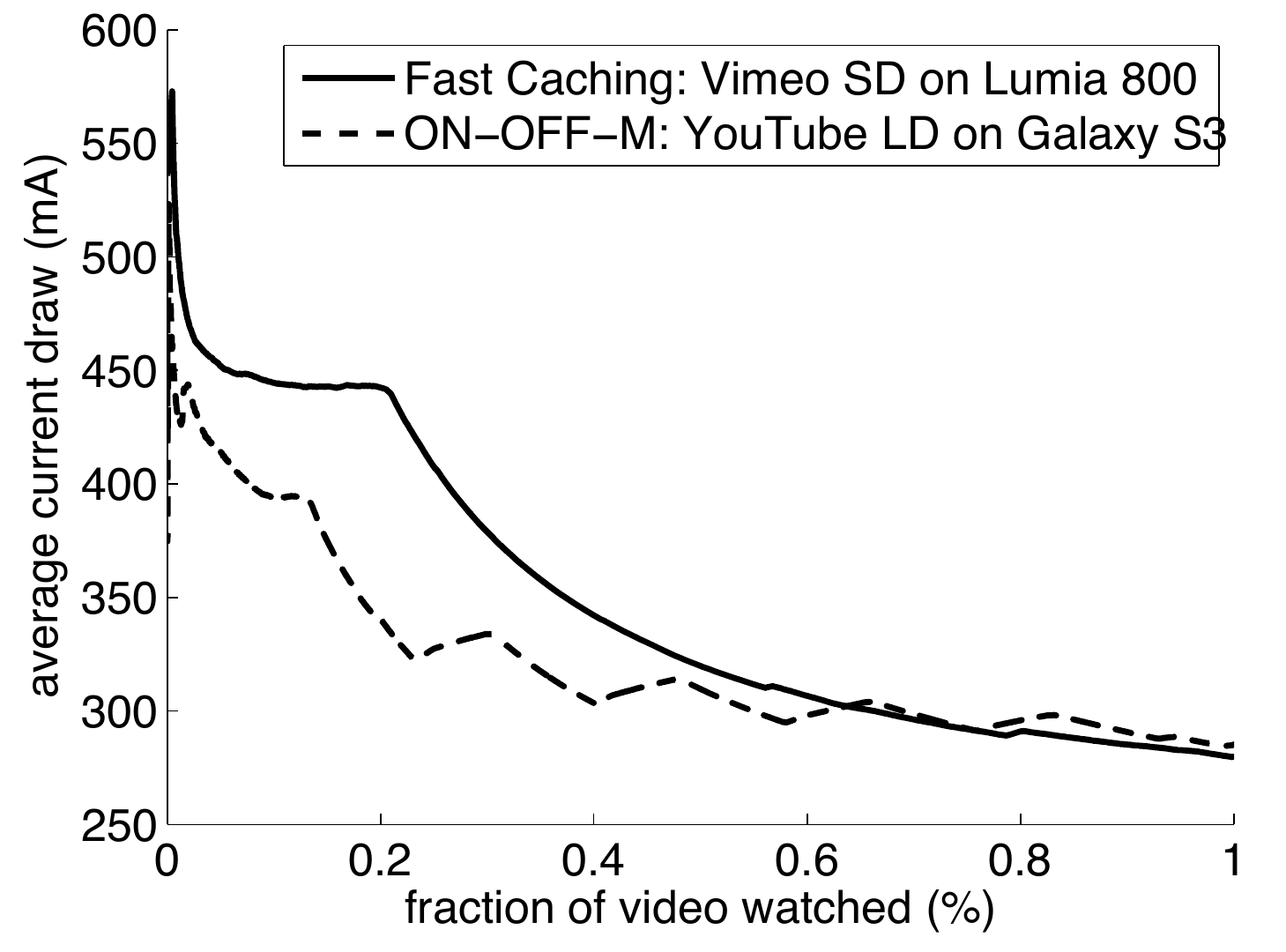}
\caption{Average draw of current as a function of viewing time for HSPA access.}
\label{fig:interrupt}
\end{figure}

From the energy consumption point of view, the downloading energy is also wasted to retrieve the unwanted content. In Figure~\ref{fig:interrupt}, we plot the average current draw for fast caching and ON-OFF-M techniques as a function of percentage of watched video computed out of the complete power traces. We see that abandoning the video watching early on would cause a hefty penalty in terms of wasted energy in both cases but the penalty gets smaller faster with the ON-OFF-M streaming making it a more attractive technique, since it is common not to watch the video completely.

\begin{figure}[!t]
\centering
\includegraphics[scale=0.48]{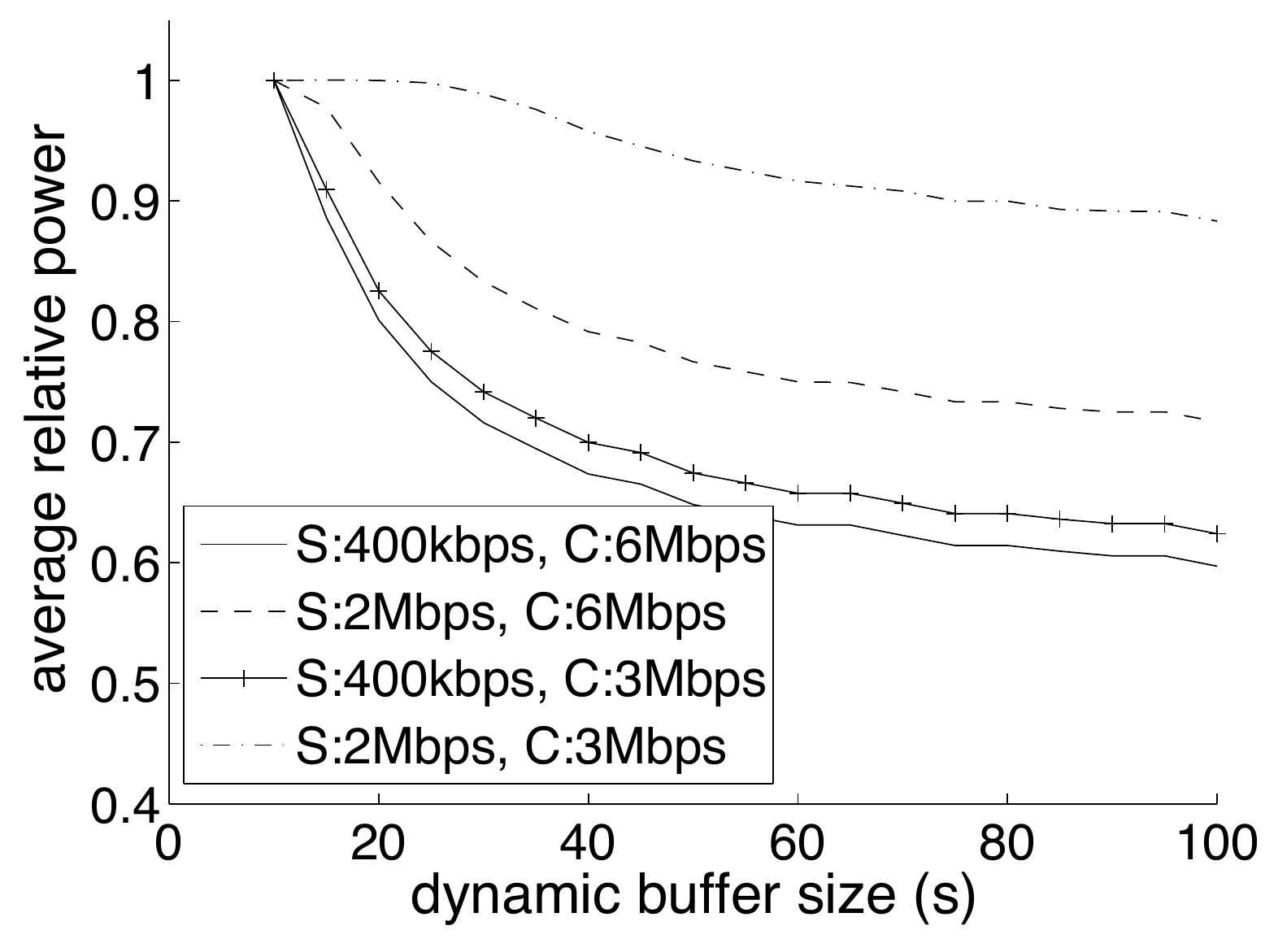}
\caption{Relative power draw as a function of dynamic buffer size for HSPA access. S is the stream encoding rate and C is the available bandwidth to download content.}
\label{fig:buffersize_vs_power}
\end{figure}

Since ON-OFF-M is the balanced technique in providing both less data waste and less energy consumption, a tradeoff between the buffer thresholds and energy consumption must be understood. Assuming that the upper threshold is fixed, i.e. the player allocates a fixed amount of memory for the playback buffer in the beginning of a streaming session, the lower threshold determines how large chunks of content will be downloaded at a time, i.e. what is the duration and frequency of the ON periods. The lower the threshold, the less frequent are the buffer refill events (ON periods), and the less power is consumed on the average. On the other hand, the lower the lower threshold is set, the higher is also the chance that there is a playback pause event when the buffer refilling begins in case a transient period of low bandwidth happens to coincide. For this reason, there is a tradeoff between risking a buffer underrun event and the power consumption which is controlled by the lower buffer threshold.

We plot in Figure \ref{fig:buffersize_vs_power} the average power draw as a function of the dynamic buffer size. The dynamic buffer size is directly determined by the lower threshold if we keep the upper threshold fixed. We notice that if there is plenty of spare bandwidth available compared to the stream encoding rate, then the buffer size should be set at least to a value around 40-50s, but setting the buffer to a larger value than that no longer reduces the power consumption significantly.

The current YouTube players in Android that use the ON-OFF-M strategy set the upper threshold to a value equalling $100s\times r_s$ and the lower one to $40s\times r_s$ where $r_s$ is the average encoding rate. These thresholds translate to a $60s$ dynamic buffer size which, in light of Figure \ref{fig:buffersize_vs_power}, strikes a good balance. Those players using ON-OFF-S technique in newer versions of Android use a 20MB buffer size. Assuming a lower threshold at zero, the dynamic buffer size would translate to 400s and 80s for videos having encoding rate of 400 kbps and 2 Mbps, respectively. With the higher quality video, the lower threshold could be set to $30-40s\times r_s$ in order to safeguard from buffer underrun events, and that configuration would still provide good energy efficiency when using HSPA.

\section{Related Work}
\label{seven}

The diverse nature of existing popular mobile streaming services in delivering better user experience, and the resulting energy consumption characteristics have so far not been completely uncovered. Krishnan et al.~\cite{Krishnan:2012} studied the effect of initial joining time and playback pause events on the engagement in watching videos for fixed host users. Their findings were such that users cannot tolerate more than 2 seconds of joining delay and if a pause event persists more than 1\% of total duration of the video the engagement decreases. Balachandran et al.~\cite{Balachandran:2013} proposed a machine learning approach which tries to improve the engagement further by selecting the appropriate CDN according to the bit rate of the content.

Many papers have studied the energy efficiency of multimedia streaming over Wi-Fi and developed custom protocols or scheduling mechanisms to optimize the behavior. Examples of such work range from proxy based traffic shaping and scheduling to traffic prediction and adaptive buffer management~\cite{hoque12survey}. However, streaming over HSPA and the specific nature of the streaming services and client apps provide new challenges that these solutions cannot overcome. Balasubramanian et al.~\cite{balasubramanian09imc3g} studied 3G power characteristics in general and quantified the so called tail energy concept.



The most popular streaming services, especially YouTube, have been subject to numerous measurement studies in recent few years. Xiao et al.~\cite{Xiao:youtube} measured the energy consumption of different Symbian based Nokia devices while using a YouTube application over both Wi-Fi and 3G access. A similar study was done by Trestian et al.~\cite{Trestian} for Android platform. They investigated energy consumption while streaming over Wi-Fi at different network conditions and studied the effect of video quality on energy consumption. However, these studies did not consider the details of traffic patterns and their impact on the energy consumption.

In a measurement study, Rao et al.~\cite{Rao:2011} studied YouTube and Netflix traffic to different smartphones (iOS and Android) and web browsers accessed via Wi-Fi interface. They found three different traffic patterns of YouTube. In a similar passive measurement study, Finamore et al.~\cite{Finamore:2011} also analyzed YouTube traffic to PCs and iOS devices accessed via Wi-Fi and demonstrated that iPhone and iPad employ chunk based streaming. Qian et al.~\cite{29.feng} explored RRC state machine settings in terms of inactivity timers using real network traces from different operators and proposed a traffic shaping solution for YouTube which closely resembles the ON-OFF streaming technique. 

Liu et al.~\cite{liu11mm} studied power consumption of different streaming services. However, the scope of their study is considerably different from ours. They limit their study to streaming over Wi-Fi and performed experiments with only iPod, while we explored all the major mobile platforms and contrasted Wi-Fi and HSPA energy consumption in~\cite{hoque2013wowmom}.

In contrast to these studies, our contributions are the followings. (i) We investigated the traffic pattern of the streaming techniques and the characteristics which influence the choice of a streaming technique. (ii) We measured the initial joining time that varies according to the service, quality of the content and wireless access. (iii) We examined the playback buffer status of the players during playback to understand to which extent they can avoid a playback pause event in case of spurious network condition. (iv) We also studied the impact of the streaming techniques on the energy consumption on different smartphones using Wi-Fi, HSPA and LTE. (v) Finally, we proposed playback buffer configurations for ON-OFF mechanism, which can ensure significant energy savings, reduce data waste, and can tolerate bandwidth fluctuations to some moderate extent.

\section{Conclusions}

We analyzed the performance of four video services in tolerating bandwidth fluctuation and the energy consumption of smartphones. Based on he measurements with the latest smartphones, we identified five different streaming techniques. The used technique depends on the service, client device or mobile platform, player type, and video quality. In general, most of the techniques are efficient in tolerating short term and long term bandwidth fluctuations by prefetching content. Since an interrupted video session can result in significant data and energy waste, ON-OFF-M provides a balance between quality of experience, and data or energy waste. We investigated how the buffer underrun and energy consumption are related and showed the optimal buffer threshold configurations with which a player can tolerate bandwidth fluctuation for 30 s to one minute, at the same time reducing data waste and saving energy.

\bibliographystyle{elsarticle-num}
\bibliography{sigproc,refs_matti}
\end{document}